\documentclass[acmsmall]{acmart}
\usepackage{csquotes}
\usepackage{caption}
\usepackage{etoolbox}
\usepackage{dirtytalk}
\usepackage{tabularx}
\usepackage{multirow}
\usepackage[export]{adjustbox}
\usepackage{tabularray}
\UseTblrLibrary{booktabs}
\AtBeginEnvironment{displayquote}{\itshape}
\AtBeginEnvironment{say}{\itshape}
\usepackage{lscape} 
\usepackage{xcolor}
\definecolor{maroonshade}{HTML}{D2B9B9}
\usepackage{soul}
\AtBeginDocument{%
  \providecommand\BibTeX{{%
    \normalfont B\kern-0.5em{\scshape i\kern-0.25em b}\kern-0.8em\TeX}}}

\setcopyright{acmcopyright}
\copyrightyear{2023}
\acmYear{2023}
\acmDOI{XXXXXXX.XXXXXXX}

\acmConference[Preprint, CSCW'24]{}{}{}
\acmPrice{15.00}
\acmISBN{978-1-4503-XXXX-X/18/06}

\usepackage{xcolor}
\usepackage{fontawesome}
\usepackage{soul}

\begin{document}

\title[Collaborative Dataset Specification for ML in Education]{Is a Seat at the Table Enough? Engaging Teachers and Students in Dataset Specification for ML in Education}


\author{Mei Tan}
\email{mxtan@stanford.edu}
\affiliation{%
  \institution{Stanford University}
  \country{USA}
}

\author{Hansol Lee}
\affiliation{%
  \institution{Stanford University}
  \country{USA}
}

\author{Dakuo Wang}
\affiliation{%
  \institution{Northeastern University}
  \country{USA}
}

\author{Hariharan Subramonyam}
\affiliation{%
  \institution{Stanford University}
  \country{USA}
}


\begin{abstract}
Despite the promises of ML in education, its adoption in the classroom has surfaced numerous issues regarding fairness, accountability, and transparency, as well as concerns about data privacy and student consent. A root cause of these issues is the lack of understanding of the complex dynamics of education, including teacher-student interactions, collaborative learning, and classroom environment. To overcome these challenges and fully utilize the potential of ML in education, software practitioners need to work closely with educators and students to fully understand the context of the data (the backbone of ML applications) and collaboratively define the ML data specifications. To gain a deeper understanding of such a collaborative process, we conduct ten co-design sessions with ML software practitioners, educators, and students. In the sessions, teachers and students work with ML engineers, UX designers, and legal practitioners to define dataset characteristics for a given ML application. We find that stakeholders contextualize data based on their domain and procedural knowledge, proactively design data requirements to mitigate downstream harms and data reliability concerns, and exhibit role-based collaborative strategies and contribution patterns. Further, we find that beyond a seat at the table, meaningful stakeholder participation in ML requires structured supports: defined processes for continuous iteration and co-evaluation, shared contextual data quality standards, and information scaffolds for both technical and non-technical stakeholders to traverse expertise boundaries.
\end{abstract}




\maketitle
\section{Introduction}
Education is a complex and dynamic system~\cite{koopmans2020education}. Yet, applications of machine learning (ML) in education rely on generalized approaches with narrow conceptualizations of educational knowledge to analyze learner behavior, interactions, and performance~\cite{perrotta2020deep}. Consequently, the adoption of ML applications in school administration, instruction, and learning has led to issues of fairness, accountability, transparency, and utility in their implications for practitioners and vulnerable student populations~\cite{baker2022algorithmic, perrotta2020deep, pedro2019artificial, holmes2022ethics, reich2017good}. Harms include systematic inequalities in recommender systems~\cite{marras2022equality} and high-stakes automated decision-making~\cite{aied_chapter}, surveillance and civil rights concerns in facial recognition systems~\cite{whittaker2018ai}, data privacy concerns~\cite{potgieter2020privacy, feathers-1, feathers-2}, and disparities in student propensities to consent~\cite{li2022disparities}.

These issues are rooted in the underlying challenge facing ML design and development, which include undefined policy-level guidelines~\cite{schiff2022education}, insufficient teacher education and involvement in ML development~\cite{bogina2022educating, schiff2021out, roll2016evolution}, the underdevelopment of inclusive and high-quality data systems~\cite{Ocumpaugh2014, gardner2019evaluating}, and the lack of ethical regulation and transparency in data collection, use, and dissemination~\cite{pedro2019artificial}. Traditional ML development processes undervalue the critical role of trustworthy training data and dataset accountability and largely assume data as given~\cite{10.1145/3411764.3445518}. 

Further, current engineering processes limit engagement with domain experts and end-users such as educators and students and miss important contextual features of real-world data~\cite{10.1145/3491102.3517537}. When included, domain experts only converge in ML development after crucial data-related decisions have been made~\cite{subramonyam2022solving}. While researchers have created guidelines for downstream data evaluation and documentation (e.g., Datasheets for Datasets~\cite{gebru2021datasheets}), standard practices remain undefined in upstream data specification~\cite{heger2022understanding}. Resolving these issues requires addressing the tensions between ML innovations, engineering priorities, and teacher and student needs. Concretely, to create ethical and human-centered ML experiences for education scenarios, we need early collaboration between educators, students, and ML practitioners.

The recent methodological shift in ML practice to re-prioritize the design and quality of training data (i.e., data-centric AI) presents an opportunity to involve teachers and students early in the design of the ML data pipeline~\cite{subramonyam2021can}. However, mere participation is \textit{not enough}~\cite{sloane2020participation}. To truly involve teachers and students, we argue they must be provided with the necessary training and resources to understand and contribute to the design process. This includes ensuring that their input and feedback are considered, working with them to resolve knowledge gaps, contextualizing data needs within domain needs, and negotiating trade-offs around scope and generalizability. Further ML software practitioners should revise their work practices to prioritize domain knowledge and collaboration with domain experts.

In this work, we investigate whether and how teachers and students can work with ML practitioners to define data requirements (the backbone of machine learning models) from the ground up. While prior research has focused on co-designing ML applications with teachers (e.g.,~\cite{holstein2019co}), our work looks at the collaborative specification of dataset attributes, labels, and data collection pipeline (i.e., items in the Datasheets for Datasets~\cite{gebru2021datasheets}). We ask the following research questions: 
\begin{itemize}   
    \item  \textbf{RQ1:} What do diverse stakeholders bring to the table when co-designing data specifications?
    \item \textbf{RQ2:} How can we systematically support and amplify diverse stakeholder voices in the ML data specification process?
\end{itemize}

To investigate these research questions, we conducted a series of co-design sessions engaging experts and stakeholders across domains in collaborative data specification for ML applications in education. Forty participants took part in our study, representing ML engineers, teachers, students, UX designers, and legal experts roles. During these sessions, stakeholders defined dataset characteristics, discussed representativeness and validation criteria, developed labels, and planned ethical data collection strategies for several common application scenarios \cite{niemi2023ai} (e.g., student drop-out risk prediction, automated essay grading, student engagement image classification).  We find that teachers and students play a crucial role in contextualizing upstream data-related decisions in downstream use and support the identification of potential biases and reliability threats during data collection and labeling. Further, we identify challenges and needs to deepen stakeholder collaboration to ensure productive participation.

In summary, our work contributes to the emerging practice of data-centric AI, collaborative processes in human-centered AI, and the growing literature on practitioner needs regarding ML applications in education. Through our co-design sessions, we highlight the affordances and limitations of having a seat at the table and discuss directions for future research designing collaborative processes for engaging stakeholders in the education domain. We also discuss the implications of our findings, including developing shared standards, information scaffolds, and supportive tooling to support multi-stakeholder contribution to ML data specification and evaluation. 
\section{Related Work}
The potential for ML systems to create or exacerbate biases, unfairness, and downstream ethical harms has received academic attention across disciplines. The focus of this work investigates the engineering processes in the research and industry environments that build these ML systems. Prior work has highlighted an urgent need for the organizational adoption of tooling and internal processes that support the responsible development and maintenance of fairer systems~\cite{latonero2017tech,sculley2015hidden}. These calls to action emphasize two high-level practices: focusing on data work and involving context in the design and development of ML applications \cite{leslie2022data}. Here we first synthesize existing literature on data practices in ML and then situate our work in current approaches to data documentation and stakeholder collaboration.

\subsection{ML Data Pipeline and Current Practices}
Compared to software application development, machine learning applications require the complexities of discovering and managing data~\cite{amershi2019software}. The machine learning lifecycle begins with data management, underpinned by a set of system-level requirements, which produces the training dataset used to drive the model learning, model verification, and model deployment stages of the ML workflow~\cite{ashmore2021assuring,guo2013data}. Data management consists of multiple steps, including data acquisition, data annotation, data pre-processing, data augmentation, and data validation~\cite{akkiraju2020characterizing}. During data acquisition, collecting examples may take the form of searching and indexing existing datasets, distorting and deriving synthetic examples from existing datasets, or creating datasets through data generation techniques \cite{whang2023data}. During data annotation, labeling examples may involve the utilization of existing labels, or manually or automatically generating new labels~\cite{whang2023data}. The data pipeline additionally encompasses the devices and processes involved in storing and moving data~\cite{munappy2020data}. The creation of data used to develop ML systems often requires costly manual work but this work critically affects the trustworthiness of the resulting model~\cite{liang2022advances}. 

Despite the complexity and significance of data management, current industry practices rely on model-centric development, in which engineering resources are dedicated primarily to iterating the model architecture or training procedure to improve the benchmark performance~\cite{liang2022advances}. Prior work has found ‘discretionary’ practices in system design~\cite{10.1145/3411764.3445518, passi2020making}, ambiguous roles and responsibilities within teams~\cite{saltz2017ambiguity}, and reliance on individual engineers to identify issues and address ethical concerns~\cite{rakova2021responsible}. Furthermore, traditional engineering processes limit engagement with domain experts and end-users, separating the work of technical development and understanding end-user requirements~\cite{10.1145/3491102.3517537}, and prioritizing technical affordances over the problems of practitioners and real-world contexts~\cite{kerner2020too, birhane2021values}.  These ad hoc and technology-focused engineering practices have resulted in haphazard data management, in which decisions regarding the definitions of data are forgotten beneath a series of additional decisions, opportunities, improvisations, and assumptions~\cite{muller2022forgetting}. Practitioners developing ML systems currently face challenges across multiple steps of the data pipeline, including finding, understanding, preparing, and validating data~\cite{polyzotis2018data, polyzotis2017data}. Audits of dataset development work have found practices that value efficiency over care~\cite{paullada2021data}, resulting in an overwhelming majority of datasets that do not meet quality standards~\cite{nagle2017only}. Data-centric practices are undervalued in conventional ML development, resulting in compounding downstream negative effects ~\cite{10.1145/3411764.3445518, richardson2019dirty}.

\subsection{Data-Centric AI and Data Documentation}

To address the limitations of model-centric AI practices, recent work has started to focus on data-centric practices, producing supportive tooling for maintaining data repositories and facilitating data annotation and validation~\cite{liang2022advances}. Research in data-centric AI has primarily addressed the downstream harms of low-quality data through the creation of numerous frameworks for facilitating data accountability and transparency through clear documentation practices~\cite{gebru2021datasheets, diaz2022crowdworksheets, richards2021human, arnold2019factsheets, bender2018data, chmielinski2022dataset, pushkarna2022data,wang2022documentation}. The dataset documentation literature introduces standardized processes for datasets to be accompanied by information identifying their motivation, context, composition, features, collection process, biases, recommended uses, and so on (e.g., DataSheets)~\cite{gebru2021datasheets}. Documentation frameworks help engineers understand ethical issues in training data~\cite{boyd2021datasheets} and provide important guidelines supporting accountability in data quality standards. 

However, prioritizing data work also necessitates supporting the collection and curation of high-quality data sets in the first place~\cite{holstein2019improving} and addressing the upstream work of defining dataset requirements~\cite{10.1145/3411764.3445518}. Ideal data-centric practices begin with specification and defining data requirements according to application needs, but ML systems commonly suffer from incomplete or misinterpreted requirements~\cite{challa2020faulty, d2020underspecification, hullman2022worst, akkiraju2020characterizing}. Practices that support the specification of dataset requirements early in the data pipeline are understudied in the data-centric ML literature. In education settings, appropriate data specification design in the early stages of ML development is key to mitigating ethical harms in a high-stakes domain~\cite{baker_algorithmic_2022, aied_chapter}. Prior work evaluating AI fairness in education has encouraged research to interrogate the definition of ML problems and data collection procedures and evaluate the quality of training data~\cite{aied_chapter}. Our research investigates the proactive process of data specification, anticipating the evaluative components of documentation frameworks.

\subsection{Domain Context and Stakeholder Collaboration}
Data is inextricably bound to place and community~\cite{taylor2015data}. The context encoded in data and the context of data production is critical to understanding datasets and their downstream applications~\cite{vertesi2011value}. Placing data in their temporal, geographic, and social context, disciplinary norms, and worldly representativeness is a key component of making sense of data~\cite{koesten2021talking}. Prior work has called for incorporating more domain knowledge~\cite{veale2018fairness}, developing domain-specific performance metrics~\cite{holstein2019improving, shi2020artificial}, and creating frameworks for documenting context-specific intended use cases~\cite{chmielinski2022dataset}. 

A growing body of research has addressed the elevation of domain context through the study of collaboration and stakeholder participation. AI and HCI communities have increasingly called for more stakeholder participation in the design, development, and maintenance of ML systems~\cite{delgado2021stakeholder, leslie2022data, weber2022organizational, boyarskaya2020overcoming, tomavsev2020ai,muller2019data,kross2021orienting,zhang2020data}. However, meaningful collaborative practice is complicated by the language boundaries of domains and the power dynamics at the intersection of communities of practice.

Firstly, collaboration in social applications of ML involves the complexities of cross-discipline communication. Development practices rooted in silos of expertise limit communication between disciplines. Subramonyam et al.~\cite{subramonyam2022solving} investigated co-creation processes between engineers and user-experience designers and found a separation of concerns between engineers and domain practitioners. Technical experts explore machine learning capabilities independently while making erroneous assumptions about human behavior and contextual needs. Passi and Jackson~\cite{passi2018trust} similarly found a separation of concerns among data science and business analyst experts dividing system accountability tasks. Mao et al.~\cite{mao2019data} studied the collaborative practices between data scientists and bio-medical scientists and found that these distinct roles often struggle to establish common ground regarding research projects. Work in stakeholder collaboration has additionally emphasized the importance of translation between different forms of knowledge ~\cite{williams1993translation}. Hou et al.~\cite{hou2017hacking} studied collaborative roles between technical and non-technical workers in a civic data hackathon, noting that the different stakeholders spoke different languages. Collaboration required organizers to understand both data science and context to serve as brokers and translate needs across disciplines. Domain stakeholder involvement requires transparent and interpretable technical explanations~\cite{estivill2022constructing}, but materials for educating stakeholders on ML are scarce~\cite{bogina2022educating}. Domain experts face barriers to participation in ML development and decision-making due to persistent knowledge gaps~\cite{10.1145/3491102.3517537}.

Secondly, the creative cooperation between stakeholders and technology designers requires the negotiation of values across communities of practice. The involvement of stakeholders in collaborative design efforts evolves from a tradition of participatory design, which highlights an agenda of democratizing innovation by shifting existing power structures and creating a hybrid space between the domains of technology designers and impacted users \mbox{\cite{10.1162/DESI_a_00207, doi:10.1080/15710880701875068}}. Equitable and community-based participatory design emphasize methods that are sensitive to the needs and practices of communities \mbox{\cite{rosner2016out}}. They aim to foster creativity, learning, and cultural production \cite{disalvo2012participatory} to design solutions that are considered successful by community metrics \mbox{\cite{harrington2019deconstructing}}.

In the education domain, participatory design methods are rarely deployed in the development of AI tools. Though stakeholder involvement is critical for the creation of useful and socially responsible products~\cite{buddemeyer2021words}, teachers are often marginalized in technology discussions and engaged only as accessories during the implementation of ML systems~\cite{roll2016evolution}. Michos et al.~\cite{michos2020involving} collaborated with educational practitioners to understand practical challenges and iteratively evaluate solutions through workshops and implementation settings, following the structure of design-based research. Holstein et al.~\cite{holstein2019designing} involved teachers and students in “participatory speed dating” in order to solicit design feedback regarding AI applications in education. Other studies in education involve end-users through need-finding interviews and product design feedback~\cite{zhang2022storybuddy, zhou2021investigating}. While such consultations are valuable, teachers and students are often engaged in a limited capacity as end-users. By participating only in later stages of ML development, long after crucial data-related decisions have been made, opportunities for envisioning equitable design solutions are limited.

In the ML pipeline, data specification is a unique high-leverage stage of involvement for stakeholder participation. End users and domain experts may play a critical role in making transparent what is valued in the data~\cite{10.1145/3442188.3445918}. When engaged early, multi-stakeholder involvement may contribute significant insights to the design of collection and labeling procedures, validation and evaluation measures, modeling choices, and downstream use and maintenance of the dataset and application.  By involving diverse stakeholders in education, our work further investigates the expanded role and contribution of teachers, students, engineers, designers and legal professionals in the co-design of data specifications. We position our work at the understudied intersection of stakeholder collaboration in the design of ML data specifications, situated in the unique and high-stakes context of education.
\section{Methodology}

To investigate collaborative interactions between stakeholders, we conducted structured co-design workshops with engineers, designers, legal professionals, domain experts, and data subjects (i.e., individuals whose data will be collected). In each workshop, we presented participants with a potential application of ML in the education domain and asked them to collaboratively generate the data specifications for the ML model. The workshop sessions were held virtually via Zoom, with one individual representing each stakeholder's role (a total of 10 workshop sessions). Each workshop session lasted $120$ minutes. Our institution's IRB approved the study. Participation was voluntary, and all participants were compensated with \$50 for their involvement.

\subsection{Participants}
We aimed to recruit one participant from each of the five roles for each session. Because our study is anchored in the education domain, we involved educators in the domain expert role and students in the data subject role. Aside from the student role, all other roles required participants to have at least one year of relevant professional experience. We recruited participants through direct email, mailing lists at university departments and technology companies, and social media posts shared by groups involved in the intersection of AI, ethics, and design. As shown in Table~\ref{table:participants}, all but one session had participants from four of the five roles. Participants with expertise in the legal and ethical AI domains were challenging to recruit as it is an emerging role in practice. However, in developing the study protocol, we consulted with a legal AI scholar to provide adequate guidance for the group in thinking about legal and ethical requirements and constraints. Further, in cases where two or more scheduled participants were absent, we rescheduled the session and compensated those who were present with an additional \$20 for their time (a total of 3 sessions). For session 10, we decided to proceed with the session with three participants as legal professionals were challenging to recruit and schedule. In total, we conducted workshop sessions with $40$ participants.

\begin{table}[h!]
  \begin{center}
    \begin{tabular}{l|l|l} 
      \textbf{Session} & \textbf{Design Scenario} & \textbf{Participants (Years of Experience)}\\
      \midrule
      1 & Student Engagement Image Classification & E (25 yrs), T (18 yrs), S, D (2 yrs)\\
      2 & Student Engagement Image Classification & E (3 yrs), T (30 yrs), S, D (2 yrs)\\
      3 & Student Engagement Image Classification & E (15 yrs), T (2 yrs), S, L (7 yrs)\\
      4 & Resume-based Career Recommendation & E (5 yrs), T (9 yrs), S, D (1 yrs)\\
      5 & Student Drop-out Risk Prediction & E (3 yrs), T (3 yrs), S, D (5 yrs)\\
      6 & Student Drop-out Risk Prediction & E (3 yrs), T (3 yrs), S, D (1 yrs)\\
      7 & Student Drop-out Risk Prediction & E (3 yrs), T (8 yrs), S, D (2 yrs), L (5 yrs)\\
      8 & Automated Essay Grading & E (2 yrs), T (5 yrs), S, D (1 yrs)\\
      9 & Automated Essay Grading & E (7 yrs), T (7 yrs), S, D (2 yrs)\\
      10 & Automated Essay Grading & E (1 yrs), T (3 yrs), L (2 yrs)\\
    \end{tabular}
    \caption{Each workshop session is listed with participants by role (E = machine learning engineer, T = teacher, S = student, D = designer, L = legal/ethics professional) and the associated design scenario. Years of experience in their fields of expertise are indicated parenthetically for each professional stakeholder.} 
\label{table:participants}
  \end{center}
\end{table}

\subsection{Workshop Protocol}
\begin{figure*}[t!]
\includegraphics[width=\textwidth]{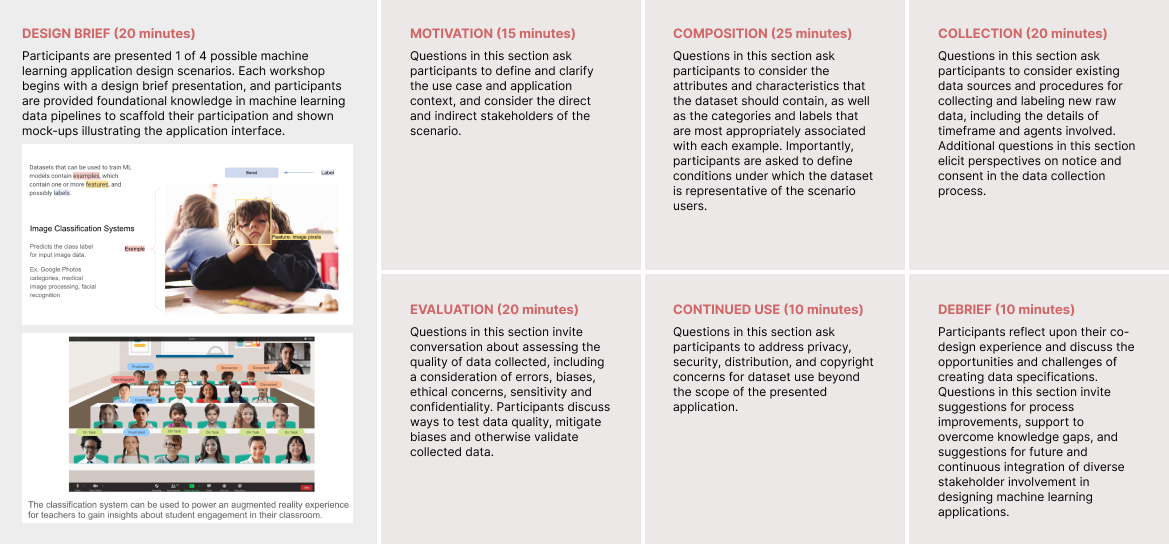}
\caption{Overview of our study protocol, including a design brief and high-level objectives for each of the data co-design sections.}
\label{fig:protocol}
\end{figure*}

Motivated by prior research studying collaborative AI design~\cite{subramonyam2021towards}, we opted to anchor our workshops on concrete applications of AI in education. Further, we used current \textit{guidelines} on human-centered data specifications and desiderata about \textit{data documentation} as a starting point to develop our workshop protocol. Concretely, the first and second authors analyzed the topics and questions in Datasheets for Datasets~\cite{gebru2021datasheets} to identify those questions that could benefit from multi-stakeholder inputs and can be \textit{proactively} specified before actual data collection. For instance, questions about attributes of each data instance, the meaning of representativeness for the dataset, and collection procedures can all be described upfront. In contrast, questions about sample size and data split between training and test sets are better defined in the later stages of the ML pipeline. As shown in Figure~\ref{fig:protocol}, the questions correspond to five main topics for our workshop protocol, including (1) Motivation, (2) Composition, (3) Collection, (4) Evaluation, and (5) Continued Use. Further, to support discussions around each set of questions, we developed guiding prompts and examples based on human-centered data guidelines~\cite{pair2019}.

To establish goals and a common language, workshops began with the presentation of a machine learning design scenario in education and a high-level explanation of the role of data in the intended application. Next, we provided participants with a data specification document with questions detailing considerations in each stage of the data pipeline. While the group collectively brainstormed ideas verbally, a research coordinator facilitated the session and recorded points of consensus in the data specification document screen-shared on Zoom. Our goal was to encourage discussions regarding priorities, trade-offs, and ethical concerns across diverse stakeholders to define data requirements. After each workshop, we asked participants a series of reflection questions to understand challenges in the co-design process and for additional tools or support to improve collaboration. We recorded each workshop session and collected generated data specification artifacts. Below, we detail the main steps in our protocol.

\subsubsection{Design Brief}
We prepared four ML application design scenarios for anchoring the workshop data specification activity. The scenarios were inspired by recent popular applications of machine learning in education, and our selection favored scenarios using different forms of input data. We aimed to understand whether and how different data types supported and challenged collaboration. The scenarios included a student engagement classification system using \textit{image} data, a student dropout early warning system using \textit{tabular} academic record data, and an automatic essay grading system using \textit{text} input data. Initially, we intended to use resume-based career recommendations as an application of text data in ML. However, based on feedback from session 4, we observed that K-12 educators were less familiar with the application use case. Hence, for the remaining sessions, we opted for AI-based essay grading, which is more familiar to K-12 teachers. In each scenario presentation, workshop participants were shown examples of inputs and outputs to the model, as well as a visual mock-up of the use case and application interface. We broadly described the training data expected through an illustration of the data pipeline with sample features and labels, while noting the many uncertainties in the data specifications left for participants to consider~\cite{aimeetsdesign}. Workshop participants were then presented with a data specification document and the co-design task. We explained the data specification document as a guidebook for software teams to collect and evaluate the training data used to build the ML application specified in the design scenario. We additionally emphasized to participants that they should work collaboratively, seek the perspectives and expertise of one another, and lean into their unique stakeholder roles to make design decisions. Our study materials are included as a supplement to the paper. 

\subsubsection{Motivation} 
In the first stage of the protocol, participants were asked to define the details of the use case and application context for the ML application described in the design brief. Questions in this stage asked stakeholders to identify the people who will directly interact with the system, be directly impacted by the system's operation, or could have a stake in how the system is created, used, or managed. We provided guidance in defining direct and indirect stakeholders and prompted participants to specify the characteristics of the relevant users and environments. Finally, we invited participants to brainstorm and elaborate concrete scenarios to situate the design task and build common ground to support subsequent steps in the protocol.

\subsubsection{Composition}
In the composition stage of the protocol, participants were first asked to consider the attributes, characteristics, and example instances that the training dataset should contain. We prompted groups that the dataset could contain multiple types or mediums of examples (e.g., documents, photos, people, countries, etc.), and we suggested that groups engage in a generative process with an eye to features that would be predictive of the target ML scenario outcomes (e.g., \textit{"which factors do you think could contribute to whether a student might be at risk of dropping out?"}). Next, participants were tasked with designing the categories and labels that are most appropriately associated with each example. We reminded groups that their chosen labeling schema would define the structure of outputs from the ML model, and we prompted participants to consider the specific context they had chosen in the previous stage. Importantly, participants were asked to define conditions under which the dataset is representative of the scenario users, including the representation and distribution of subgroups. By the conclusion of this stage, stakeholders converge on the specifications for the examples, features, labeling schema, and relative quantities of data representing important traits in the composition of the training dataset.

\subsubsection{Collection}
In the collection stage of the protocol, participants were asked to design procedures for collecting data based on specifications defined in the previous step. We prompted that approaches could involve the collection of new raw data from relevant communities or accessing and re-purposing existing data sources. We explained that data collection mechanisms could additionally include direct reports by subjects (e.g., survey responses) or inference and derivation from other data (e.g., part-of-speech tags). We additionally asked participants to consider the details of the data collection timeframe and people involved in the collection process, as well as how consent should be requested and provided by individuals represented in the dataset. Participants were then asked to design procedures for labeling the examples in the dataset, specifying the people involved (e.g., domain experts, crowd workers, students), compensation, and timeline. Finally, we invited participants to brainstorm precautions that should be taken to avoid biases introduced by the data collection process and revise the collection procedure to address these risks. At the conclusion of this stage, stakeholders converge on a set of specifications for data collection and labeling and develop an expectation for the shape of the data in preparation for the next step.

\subsubsection{Evaluation of Data Quality and Data Cleaning}
In the evaluation stage of the protocol, participants were asked to assess the quality of the data collected. To scaffold understanding of data evaluation, we asked participants to first consider how they might measure the quality of a model, noting the desirable and undesirable behaviors in the downstream system. Using the discussion of model quality as common ground, we invited participants to design metrics and validation processes to test for data quality. We prompted groups to consider the potential errors, biases, and ethical concerns in collected data, as well as the characteristics of high-quality data that would inspire confidence in training a high-quality model. Next, we asked participants to design high-level data cleaning procedures, including the removal of low-quality or erroneous examples, the handling of sensitive or confidential data, methods for addressing the under-representation of various subgroups, and the processing of missing data.

\subsubsection{Continued Use}
In the final specification stage, participants were asked to address privacy, security, distribution, and copyright concerns for dataset use beyond the scope of the presented application. We invited participants to discuss whether the dataset should be distributed for use in future applications and to consider the mechanisms and procedures for data access. We prompted groups to consider the qualities of responsible data stewardship and to brainstorm continued implications of having specified and created the dataset.

\subsubsection{Debriefing}
At the conclusion of each workshop, participants were invited to reflect upon their co-design experience and discuss the opportunities and challenges of creating data specifications. We asked participants to share the highlights and lowlights of their collaborative co-design experience, recall moments in which they encountered or overcame knowledge gaps, and make suggestions for process improvements in engaging diverse stakeholders in the design of machine learning applications. We explained that the guiding questions used in the data specification design task are an active area of investigation, and we solicited feedback on the order, clarity, and completeness of the protocol.

\subsection{Data Analysis}
The first author transcribed all workshop sessions first using a Python script with speaker diarization and then, in a second pass, manually verified the transcribed text and speaker roles against the video recordings. Next, we conducted inductive qualitative coding in Atlas.ti~\cite{atlas.ti} using a grounded theory approach \cite{strauss1990basics} beginning with in-vivo analysis. Two authors independently open-coded the same two transcripts and collaboratively developed an initial code book, resolving disagreements by consensus. The resulting codebook consists of $53$ codes. The coding scheme included references to procedural data needs (consent, labeling, cleaning, validation, etc.), contextual data needs (representation, bias mitigation, trade-offs, etc.), and collaborative processes (translation, sharing domain expertise, making assumptions, misconceptions, etc.). Using this codebook, we coded the remaining transcripts. The first author applied the code book to analyze the remaining transcripts \cite{denzin2023sage}. Throughout the coding process, the authors wrote reflective memos describing insights and emerging themes and making connections across workshop sessions~\cite{birks2008memoing}. Once coding was complete, the research team engaged in multiple discussion sessions. In these sessions, we grouped codes and discussed memos through an iterative sense-making process to identify higher-level themes and synthesize findings across transcripts. Analyses and discussion of themes were informed by the authors’ experiences conducting the workshops, as well as by artifacts and notes produced in each session. Our analyses offer insights into the collaborative process for human-centered data specification across stakeholder domains of expertise.

\subsection{Positionality}
We acknowledge that our research perspectives and approaches are shaped by our own experiences and positionality. Specifically, we are researchers living and working in the U.S., with teaching experience and experiences working with school teachers and district personnel on technology integration, researching the fairness of AI in education, and working with AI practitioners on projects related to human-centered design. In addition, we come from a mix of disciplinary backgrounds, including Computer Science, Learning Sciences and Technology, Education, and HCI, which we have drawn on to conduct prior research into sociotechnical approaches to human-centered AI design practices.  

\section{Findings}
In each workshop session, we provided a diverse group of stakeholders with a specific application of ML in the education domain. We asked them to co-design specifications for each stage of the ML data pipeline. Teams engaged in rich discussions to define dataset composition, collection and labeling procedures, and evaluation metrics. By engaging in generative design thinking, participants shared domain expertise and personal (experiential) perspectives to anticipate challenges and navigate ethical considerations for data subjects and end-users. Across all sessions, knowledge sharing and constant co-evaluation facilitated the conceptualization of a human-centered ML data pipeline from the ground up. We summarize our study findings in terms of (1) contextualizing upstream tasks with downstream use, (2) collaboration strategies across expertise boundaries, and (3) shifting roles, identities, and support needs.

\subsection{Contextualizing Upstream ML Tasks within Downstream Use}
Typical ML data pipelines are linear and comprised of distinct data and modeling tasks. Our protocol based on current data documentation templates also followed a linear organization. However, participants tended to approach specifications for each component by considering its \textit{interactions} with other stages in the data lifecycle. While current stages in the ML data pipeline are meaningful to engineering tasks, cross-discipline negotiation of concerns transcended discrete steps in the ML data pipeline. As summarized in Table ~\ref{tab:downstream_context}, domain experts across all sessions contextualized upstream ML data tasks by considering downstream application context and hypothesized the consequences of collection and modeling decisions in downstream usage.  In engaging diverse stakeholders in designing data needs in each stage of the pipeline, we find that collaborative practices can disrupt the backward-looking engineering process of retroactively improving models when performance, utility, or ethical issues surface. Here we present observations about how stakeholders \textit{proactively} anticipate challenges, consider trade-offs, recognize data unknowns, and address bias and reliability threats.

\NewTblrTheme{fancy}{
 \SetTblrStyle{caption}{\normalsize}
}
\scriptsize
\begin{longtblr}[
    theme = fancy,
    caption={Summary of stakeholders' downstream considerations in the education domain associated with upstream data specification tasks and challenges faced by teachers and students in our design workshops.},
    label={tab:downstream_context}
]{
    width=\linewidth,
    colspec={ Q[1.2] Q[3.5,l,h] Q[3.5,l,h] Q[3,l,h] },
    row{1} = { font=\bfseries },
    rowhead = 1,
}
\textbf{Upstream Data Task} &
  \textbf{Domain Contexts} &
  \textbf{Concerns} &
  \textbf{Unmet Support Needs}\\
  \hline
\SetCell[c=4]{l, maroonshade}{\textit{Composition}} \\\\
\SetCell[r=2]{c}\rotatebox[origin=c]{90}{Identifying relevant variables} &
  Training data should account for differences across \textbf{diverse educational environments} (e.g., public and private institutions, geographic location, grade level, subject of study, and mode of instruction). Representation of subgroups is required along demographic dimensions (e.g., race, gender, socio-economic status) as well as \textbf{individual learning needs} (e.g., language proficiencies, neurodiversity, disabilities). &
  Teachers and students are unsure about the feasibility and ethics of obtaining \textbf{sensitive information} (e.g., student perceptions on their relationships with their teachers). Both domain stakeholders express concern about the \textbf{fairness and utility} of the model given the numerous critical \textbf{factors that data cannot capture} about the student experience (e.g., administrative data does not indicate whether a student is experiencing homelessness or traumas outside of school). &
  Non-technical stakeholders lack technical knowledge about \textbf{data use across the ML pipeline}, including the relationship between training data, application data, and data used for validation (e.g., specifying variables for training data that may be infeasible to collect continuously in application data, hesitating to collect demographic variables under the assumption that they must be model inputs).
   \\
 &
  Nuanced \textbf{contextual interpretations} of administrative variables (e.g., separating general absences from excused absences that involve medical leave, student self-perceptions of aptitude detectable from course selection), \textbf{student out-of-school factors} (e.g., family and community support, extracurriculars, social network), and \textbf{self-reported perceptions} (e.g., writing confidence, classroom trust and safety, boredom) are impactful predictors. &
  Teachers worry about \textbf{misinterpretation of causation} as users attempt to make sense of model inputs and outputs and take subsequent \textbf{misinformed action} (e.g., administration blaming student drop-out on teaching quality despite imperfect measures, students learning to insert complex vocabulary rather than improving writing holistically). &
  Non-technical stakeholders struggle to conceptualize \textbf{how variables influence prediction}. This knowledge gap is further complicated by technical handling of different types of data and modeling choices that influence \textbf{explainability}.
   \\\\
   \hline
   \\
\SetCell[r=3]{c}\rotatebox[origin=c]{90}{Developing labeling schema} &
  \SetCell[r=2]{l}{Labels and attributes should align to \textbf{pedagogical goals} (e.g., standards-aligned rubric for essay evaluation along multiple dimensions rather than holistic scoring) and signal actions toward \textbf{improving teaching practices} (e.g., identifying lesson activities with low-engagement rather than students who seem bored, identifying specific supports required by students rather than risk of drop-out).} &
  Teachers raise concerns about the complexity of administrative and professional development efforts required to specify \textbf{followup action} and \textbf{accountability} in response to predictions. &
  \SetCell[r=3]{l}{Stakeholders lack \textbf{common ground}, leaving domain stakeholders to advocate for and explain pedagogical goals, instructional practices, organization of school systems, and sensitive issues in education.}
   \\
 &
   &
  Teachers caution against labels that cast assumptions about students and limit \textbf{student agency} (e.g., administrative repercussions from classifying students as "drop-outs",  behavior management implications from predicting student emotions). Labels impact the design of the final application and how users are trained to interact with it. &
   \\
 &
  Labeling schema should account for \textbf{multiple standards} across the education system and inherent inconsistencies (e.g., teacher discretion in grading, varying academic standards, varying state requirements for graduation). &
  Teachers worry about \textbf{academic biases} in attributes associated with quality labels in strict evaluative environments (e.g., valuing Standard American English over language familiar to students in their communities) &
   \\\\
   \hline
\SetCell[c=4]{l, maroonshade}{\textit{Collection}} \\\\
\SetCell[r=2]{c}\rotatebox[origin=c]{90}{Identifying data sources} &
  School systems maintain \textbf{administrative data} and \textbf{historical records} consisting of basic academic and demographic variables. Teachers may also be able to assist with data collection or submit data directly. &
   &
  Domain-stakeholders struggle with unknown \textbf{data ownership} and unknown data management (e.g., deferring to administration without nowing roles responsible for data management or available variables in administrative data, uncertainty about \textbf{privacy laws} or what teachers can legally share).
   \\
 &
  Educational systems include \textbf{third-party partnerships} and interactions with technology, testing, and consulting companies that privately manage data (e.g., College Board, learning management systems, national board for professional teaching). &
   &
  Stakeholders lack clarity about data collection, management, and privacy \textbf{terms from third-party systems}.
   \\\\
\SetCell[r=3]{c}\rotatebox[origin=c]{90}{Defining collection procedures} &
  Procedures must account for \textbf{legal regulations} that govern data collection in protected school-aged populations (e.g., COPPA). Collection may require multiple forms of \textbf{data use agreements} (e.g., \textbf{informed consent} from parents and legal guardians, informed consent from data subjects, data contracts with administrative data owners and organizations). &
  Opt-in consent policies may result in \textbf{sampling biases} (e.g., overhead of parental consent may deter schools or individual students from participating, volunteered essays may skew toward positive examples). &
  \SetCell[r=2]{l}{Teachers and students lack a frame of reference for what they can expect to in terms of \textbf{rights and disclosures} detailed in consent forms. Without knowing the highest standards for data privacy and security practices, they cannot evaluate the language of data agreements.}
   \\
 &
  Consent forms should build trust with \textbf{transparency} of purpose and assurances for data management, storage, sharing, and deletion. &
  Stakeholders worry about the \textbf{data privacy} implications of maintaining student identifiers and sensitive information.
   &
   \\
 &
  Procedures should account for contextual factors that may impact data quality such as \textbf{temporal variation} (e.g., differing standards and experiences at the start and end of a school year, differing activities at the start and end of a class period), \textbf{uncooperative data subjects} (e.g., unreliable or falsified student-submitted data), and \textbf{invasive collection methods} (e.g., inauthentic writing tasks, students being aware of being filmed).
   &
   &
   \\\\
   \hline
   \\
\SetCell[r=2]{c}\rotatebox[origin=c]{90}{Defining labeling procedures} &
  Labeling should emphasize student agency and fairness through incorporating \textbf{student perspectives} (e.g., allowing students to self-identify engagement). &
   &
   \\
 &
  Labelers should have \textbf{domain expertise} (e.g., experienced teachers, mental health professionals with knowledge of the age group). The \textbf{diversity} and representativeness of labelers should match that of the data subjects. &
  Teachers and students express concern about the \textbf{subjective evaluations} and biases that are unavoidable in the education domain (e.g., teacher biases in perceiving student behavior, inconsistent grading standards between teachers). &
   \\\\
   \hline
\SetCell[c=4]{l, maroonshade}{\textit{Evaluation}} \\\\
\SetCell[r=1]{c}\rotatebox[origin=c]{90}{\parbox{2.5cm}{Identifying cleaning \\and validation \\requirements}} &
  Data may be subject to \textbf{missingness} or collection limitations, including biased samples of included schools (e.g., participation from only urban charter schools) and data fields that cannot be collected (e.g., student health details). &
  Teachers worry about the \textbf{transparency of flaws} in the dataset and implications for interpreting model outputs. They note the lack of protocols for documentation and training to name the biases in the data.
   &
\end{longtblr}
\normalsize

\subsubsection{Domain Context Shapes Dataset Specifications}
In the composition stage of the protocol, participants initially approached dataset characteristics with broadly defined education contexts. However, in all sessions, teachers and students played a critical role in refining initial specifications in ways that captured the nuances of realistic downstream needs. The first section of Table~\ref{tab:downstream_context} summarizes how teachers' and students' downstream domain considerations influences their contributions to identifying relevant variables and developing labeling schema. For example, student perspectives provide insights that help to contextualize and re-interpret academic and administrative data. Though models in education commonly predict student outcomes based on an evaluation of academic achievement, student S6 explained that a feeling of academic success and well-being depends on a meaningful combination of variables:

\begin{displayquote}
S6 (Student Drop-out Risk Prediction): ``I think it's not just what their grades are. If somebody is failing out of like AP classes versus like acing like non-AP classes, I feel like, you know, the combination of those things says different stories.'' (1)
\end{displayquote}

Teachers similarly identified domain-relevant data features. While normative data practices associate diversity and demographics with a limited set of attributes, teachers highlighted the richness of what diversity means in education. For example, teachers noted the effect \textbf{environmental context} such as educational institution type, teacher experience level, urbanicity, teaching quality and subject matter, and community socioeconomic status may have on predicted outcomes.  For instance, when specifying student attentional data, teacher T2 explained: 

\begin{displayquote}
T2 (Student Engagement Image Classification): ``Diversity can mean so many different things beyond just physical attributes. It’s like diversity of environment, what they're working on, because focusing on math might look different than focusing on reading or art\ldots or if they're working with other people\ldots because those are things that could play into a student's engagement level.'' (2)
\end{displayquote}

Further, when defining subgroup representation, teachers advocated for \textbf{student-specific context} variables known to significantly differentiate learning experiences and outcomes including student age, language learner status, and first-generation or immigrant status. One contextual variable commonly raised by teachers involves the presence of learning differences or disabilities that would require an individualized education program (IEP). Teacher T1 described:

\begin{displayquote}
T1 (Student Engagement Image Classification): ``The other thing is maybe having data knowing whether a student has an IEP\ldots if a student is diagnosed with ADHD they are not going to be as focused as the student without. There are also emotional learning disabilities\ldots students who are having particular traumas at home, and these are actually pretty relevant in terms of motivation and engagement.'' (3)
\end{displayquote}

Across sessions, the range of encoded information contextualizing diversity stands in contrast to common engineering interpretations of representation standards. By sharing rich anecdotal insights and examples, teachers introduced contextual factors that prompted the group to rethink the \textbf{scope and generalizability} of dataset composition. Consequently, participants considered tradeoffs between the size of the dataset (i.e., large-scale collection of contextual variables representing diverse learners and environments ) and the scope of the application scenario. Participants expressed concerns about both the amount of data required to ensure equitable distribution of subgroups represented in the dataset and the prediction accuracy for underrepresented subgroups. Subsequently, participants considered whether the model should be designed to apply only to a specific age group, learner status, or type of school. In these discussions, engineers contextualized downstream modeling options and machine learning processes to support data composition and target application decisions.  For example, engineer E6 explained the practice of comparing multiple models, in response to a disagreement between non-technical stakeholders regarding semantic differences in public and private school data:

\begin{displayquote}
E6 (Student Drop-out Risk Prediction): ``In machine learning, it's pretty common to have multiple models and it's called ensembling where you just put them all together. You then choose the best results\ldots but then you have  different understandings of the same data, and that can be pretty useful, especially if we want to compare\ldots'' (4)
\end{displayquote}

In most sessions, engineers shared expertise highlighting the affordances of machine learning, downstream opportunities to test multiple design choices, and trade-offs in the accuracy and interpretability of technical methods. These technical contributions situate the design of dataset composition in downstream processing and modeling applications, adding methodological context to the real-world factors shared by domain practitioners. 

\subsubsection{Context Enables Identifying Bias and Reliability Threats in Data Collection}
Based on contextual knowledge introduced by teachers, all teams identified threats to data quality in the design of data collection and labeling specifications. The second section of Table~\ref{tab:downstream_context} summarizes how participants' domain contexts informed their concerns and considerations when identifying data sources, defining collection procedures, and defining labeling procedures. For example, when designing \textbf{data collection procedures}, stakeholders leaned on student perspectives as data subjects to anticipate concerns with false, malformed, and missing data. Students shared their personal experiences with survey fatigue, inadequate incentive structures, and creating fake signals. In helping the group to maximize response rates for collecting student resumes, student S4 explained their reactions to various collection strategies:

\begin{displayquote}
S4 (Resume-based Career Recommendation): ``As a student I’m probably not going to bother sending off my resume for no compensation, but compensate me and I might try and game it. However, if you do ask me for the resume with compensation and a survey, I feel like answering the survey questions is maybe going to invest me more than if I was just firing off PDFs.'' (5)
\end{displayquote}

Student S2 cautioned that observable signals from students may not align with their needs or experiences. They explained: \textit{``Students always have a way of masking. Even like on Zoom it could look like I'm looking at the screen but I'd be on my phone.''} In several sessions, stakeholders grappled with the possible impacts of uncooperative data subjects and adjusted collection procedures to prevent unwanted outcomes. Their hypothesized solutions include information campaigns to improve data subject buy-in, designing non-invasive collection methods to avoid disrupting authentic student behaviors, and enforcing the completion of data fields in the survey instrument to avoid downstream handling of missing data.

Further, data quality threats motivated stakeholders to design \textbf{data cleaning procedures} that detect and remove malformed or false data and standardize data fields. Regarding variability in the interpretation of grade labels in collected district writing samples, engineer E10 (Automated Essay Grading) explained: \textit{``Some of the schools are strict on grading, and some of them are not, so maybe we also need to align those\ldots maybe we need to do the data engineering to make all of these standardized.''} In another session, a student experience with variable grading encouraged the group to consider labeling schema derived from state standards. Student S8 explained: \textit{``To me, teachers are not always consistent with grading. They all have different perspectives on what A and B are.''} Data representativeness and appropriate distributions of diverse subgroups emerged as a recurrent theme regarding data quality in the evaluation stage. Participants designed processes to pursue this standard by returning to the design of collection procedures and augmenting data from underrepresented groups via additional collection processes, extrapolation, or borrowing data from other contexts. 
  
Third, teams brainstormed ways to account for errors in data collection during specification of \textbf{data labeling procedures.} Participants committed to hiring multiple labelers and calculating agreement scores, auditing labels in second-hand datasets, and specifying requirements for the qualifications and diversity of labelers as a precaution for obtaining reliable labels. While labeling procedures are discussed in the collection stage of the workshop protocol, participants often discussed data quality standards and data evaluation concepts to anchor their design decisions. Considering the authenticity of labelers, and drawing upon their previous experience as a special-education teacher, designer D2 suggested engaging data subjects in self-identifying labels:

\begin{displayquote}
    D2 (Student Engagement Image Classification): ``I think it would be really important actually to have students self identify. The people who are interpreting that facial expression are going to have a different interpretation than the person that made it. But like having the comparison between the students' self reflection and the teachers' perception and measuring that gap in between.'' (6)
\end{displayquote}

For sessions in which groups decided to re-purpose an existing dataset rather than collecting raw data, participants expressed skepticism over potentially biased labels and concern for the downstream effects of mislabeled examples.

\subsubsection{Validating Specifications by Mapping to Context}
Building on the newly acquired contextual knowledge, all teams referenced context as a way of assessing evolving specifications and collection procedures. The third section of Table~\ref{tab:downstream_context} summarizes how domain context shaped participants' considerations when identifying data cleaning and validation requirements. Few groups noted the importance of transparency of data processes to mitigate applied bias in downstream use cases. As an example, in the automatic essay grading scenario, one of the groups considered reusing second-hand data from a standardized testing service, while observing that work samples from students \textit{financially able} to access the service were over-represented. Specifying the communication of the biases that cannot be mitigated, teacher T8 described:

\begin{displayquote}
    T8 (Automated Essay Grading): ``I think also just naming the biases that you cannot reduce, or that you cannot address, so if you’re using a certain set of criteria that’s constructed by the AP, the emphasis on language conventions is\ldots biased towards standardized academic English. Okay, if we can’t eliminate this bias, we can at least name it in our process.'' (7)
\end{displayquote}

Further, rather than leaning on technical language and conventional engineering metrics for model performance in the evaluation stage of the protocol, domain experts encouraged groups to prioritize the evaluation of the application as a whole. Across all sessions, groups acknowledged the tradition of standardized quantitative metrics and their inability to capture the real-world effects of a machine learning application. Adding to the conversation about testing procedures validating the accuracy of the model, teacher T9 (Automated Essay Grading) advocated for the most relevant domain-specific performance metric in education: \textit{``The quality should be measured by the learning outcomes of the students\ldots like how they're responding to the feedback that they get from the tool. I know that's really hard.''} The challenge of designing evaluation specifications prompted stakeholders to revisit decisions in earlier stages of the data pipeline. This often involved returning to design decisions in the data composition stage and selecting different labels that would better support end-user goals. Despite recognizing the incompleteness of existing metrics, participants faced difficulties creating new ones that better serve contextual needs. 

To summarize the section, by moving freely along the data pipeline, participants situated data needs and machine learning processes in domain-aware contexts. They anticipated end-user experiences and proactively mitigated threats to data quality. The separation of concerns between data, modeling, and application work represents an engineering-centric framework. We find that multi-stakeholder groups engage in decision-making holistically, contextualizing data specification with use case elicitation and trade-offs in every stage of application development.

\subsection{Collaboration Strategies Across Expertise Boundaries}

While role-based knowledge boundaries have traditionally limited opportunities for collaboration between machine learning engineers and domain experts, we observed multi-stakeholder groups engaged in boundary-spanning collaborative practices. Participants employed expertise-specific strategies to overcome knowledge gaps and build cross-discipline understanding. Through practices of translation and advocacy, groups amplified diverse perspectives, built common ground, and navigated ambiguity.

\subsubsection{Translation}
Non-technical stakeholders often perceive barriers to participation in technical decision-making due to knowledge gaps in machine learning capabilities and processes. Acknowledging the lack of familiarity with ML in the education domain, teacher T9 explained:

\begin{displayquote}
    T9 (Automated Essay Grading):  ``If you were to go into a classroom, I think the majority of high school teachers in the United States\ldots if you say machine learning and natural language processing and algorithms, they have no idea what you’re talking about. That’s not because they’re stupid, it’s just because it’s a very niche topic that you don’t really hear much about when you’re in the classroom. There needs to be some sort of middle ground\ldots some kind of translation to lay folks that don’t live in this world of zero and ones.'' (8)
\end{displayquote}

While jointly negotiating data needs, engineers across all sessions facilitated collaboration through translation. Concretely, technical experts went beyond merely re-framing practitioner priorities into machine learning terms in data specifications. Engineers in the most successful co-design sessions actively shared technical knowledge to establish common ground and scaffold practical understanding for domain expert participation. One mode of translating contextual data needs into technical specifications involves \textbf{evaluating the feasibility} of teacher requests. In the data composition stage, when teacher T5 advocated for augmenting standardized exam scores with local classroom performance metrics, engineer E5 considered the feature in terms of its technical representation:

\begin{displayquote}
    T5 (Student Drop-out Risk Prediction): ``Would it be much harder to add in that layer? It’s just whether they have passed a class with a certain letter, in this case it’s a C or higher.''\\
E5 (Student Drop-out Risk Prediction): ``I wouldn’t say it's a hard feature to add, it sounds like a binary feature. It’s a yes or no, right? You add a column, and the value is yes or no. Yeah, I think that’s feasible.'' (9)
\end{displayquote}

Engineers across all sessions applied their technical knowledge to support feature requests from teachers whenever feasible. Translation occurred in the encoding of teacher-raised relevant data fields, as well as the planning of technical processes in the data collection and evaluation stages to accommodate use case concerns. Engineers engaged in translation by \textbf{clarifying machine learning processes} to support broader practitioner concerns and values. For example, given the publicity surrounding privacy violations and biases along demographic dimensions in machine learning, nontechnical stakeholders displayed a sensitivity to collecting race, age, and gender variables. Addressing confusion and unease about the collection and use of demographic data, engineer E2 explained:

\begin{displayquote}
    E2 (Student Engagement Image Classification): ``You can decide to use [demographic variables] to understand your data, and then you already know the data is potentially biased. So when you build your model, you keep that in mind, and you refine your model to cope with that bias.'' (10)
\end{displayquote}

By translating domain priorities into evaluative specifications, the engineer reached across knowledge boundaries and deepened a collective understanding of the use of demographic data in machine learning processes. As a practice, translating allowed engineers to use their technical expertise to amplify the voices of practitioners, enabling non-technical stakeholders to contribute to the construction of human-centered data needs. Using the shared scenario context, engineers \textbf{explained trade-offs} in data and modeling choices, building the technical foundation to support domain expert participation. While considering the representation of diverse school settings in the data composition stage, engineer E6 described their considerations in specifying the scope of a model:

\begin{displayquote}
    E6 (Student Drop-out Risk Prediction): ``If you train this model for just this one school, then you would be looking at all of the previous data that you have from that school\ldots with the downside being that you might not have enough data for the model to learn from, or it might draw the wrong conclusions. One of the benefits of training a model on all the schools in the district is that you have a lot more data points. But the downside of that is like maybe you're at a really poor school, and all the other schools in your district are really rich, so the drop-out patterns might be different.''\\
T6 (Student Drop-out Risk Prediction): ``They both seem to have downsides, but maybe per-district is better, because if it's generalized for everyone, then the inaccuracy is higher, but there's a lot more data, so it's better to be more accurate.'' (11)
\end{displayquote}

By leaning on the design scenario, the engineer contextualized the effects of technical decisions in domain-relevant terms, enabling the teacher to engage in the evaluation of the trade-off. While technical terms such as “accuracy” and “generalization” had been used previously in the workshop, they had not been taken up by the teacher. By translating the contextual costs and benefits of modeling choices, the engineer empowered the teacher to then take up technical language and contribute to decision-making.

In many sessions, the technical stakeholders took additional care to educate non-technical stakeholders regarding technical details through extended dialogue, actively scaffolding their uptake of technical language in the design process. In a few sessions, engineers employed metaphors and likened ML to familiar analog processes, tailoring technical knowledge in their explanations to serve as a broker between domains. Translation was practiced predominantly by technical experts. Due to the social nature of the education domain, technical experts may more easily make assumptions about educational contexts without requiring translation from domain experts, while the infusion of technology in education is a recent and disruptive change foreign to many domain practitioners.

\subsubsection{Advocacy}
The high-stakes nature of the educational field necessitates developing machine
learning applications prioritizing practitioner experiences. In the collaborative context, domain experts engaged in advocacy, leaning into \textit{extended discourse and emotion-driven language}, urging out-of-domain stakeholders to confront the complexities of education systems and hidden implications of data decisions. As student S4 described, \textit{``you have to try and be an advocate, and if you're going to deploy a system like this you're going to have to come up against people who do advocate for other interests.''} Indeed, teachers and students across sessions characterized their collaborative participation as advocacy. They advocated for fairness and utility priorities motivated by their domain contexts and lived experiences while negotiating cross-cutting requirements. 

By surfacing critical downstream implications of data labels, feature encoding, and modeling choices, teachers and students voiced values and sensitivities central to the education space. Student S5 advocated for data subjects by explaining that privacy violations and data misuse put students at risk of negatively impacting future educational opportunities.

\begin{displayquote}
    S5 (Student Drop-Out Risk Prediction): ``I would be concerned about teachers or administrators or a committee\ldots overseeing the results\ldots the degree of embarrassment if I did show up as someone likely to drop out\ldots that would imply that you know you're not performing well and something's wrong.'' (12)
\end{displayquote}

Teacher T6 similarly expressed concern about downstream harms for students due to the severity of language characterizing labels (e.g., "dropped out" and "did not drop out") in the student drop-out prediction scenario. They warned: \textit{``Then these students will be labeled like dropouts, and then it gives administrators a reason to push students like this out of school.''} Teacher T6 instead advocated for student-centric labels and reframed the application scenario to predict whether a student was \textit{``on track to graduate''}. The complex structure of educational systems produces role-based differences in interests, priorities, and interpretations of model results. While school and district administrators value drop-out metrics, teachers prefer a reversed framing featuring student progress towards positive goals, aware of the real-world implications for how students flagged by the system may be treated. By explaining their experience-motivated understanding of mentalities and practices in the downstream application context, teachers in several sessions advocated for data specifications that avoid the perpetuation of system inequalities.

\begin{figure*}[t!]
\includegraphics[width=\textwidth]{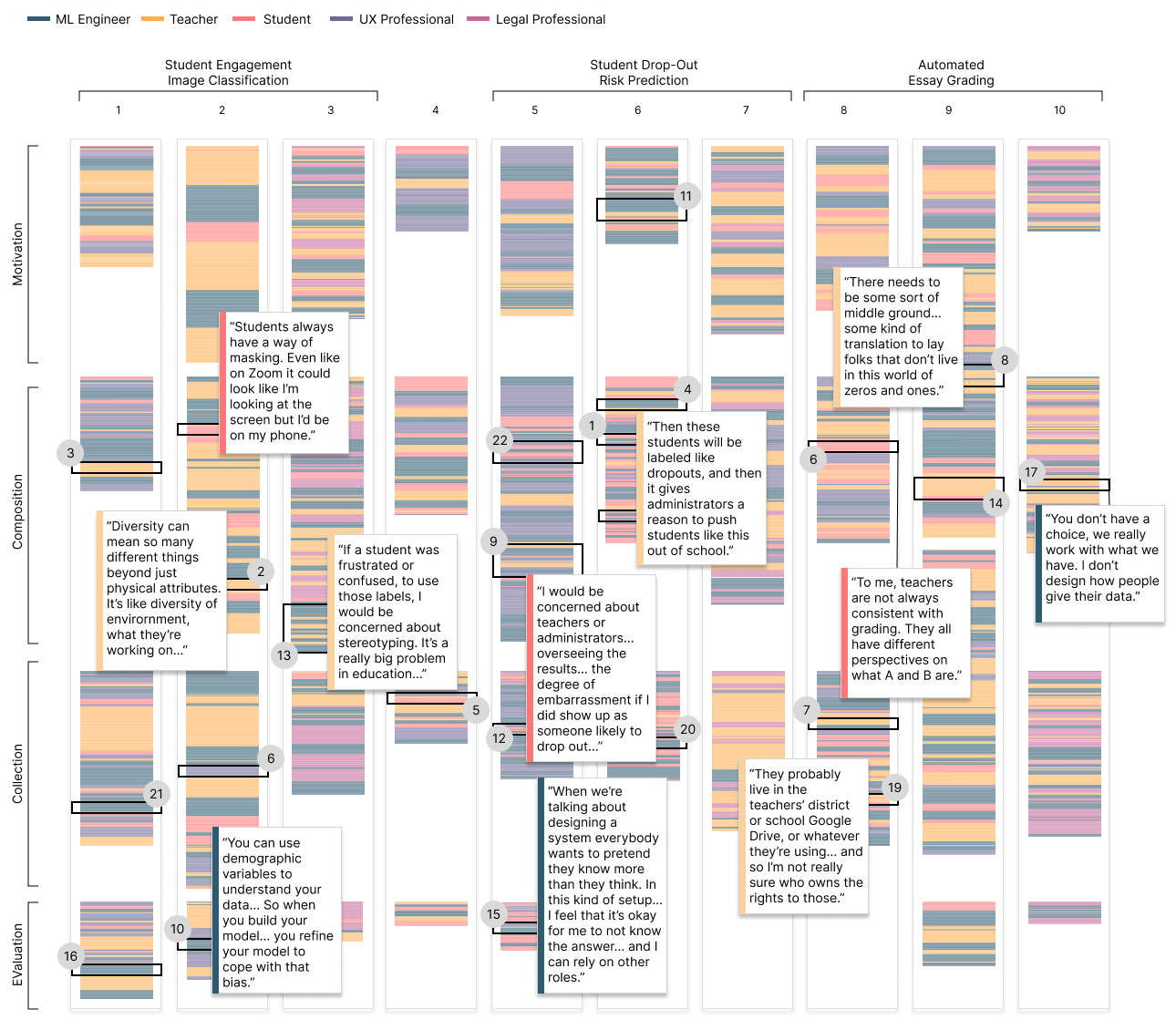}
\caption{Visualization of role-based contributions in workshop discussions across stages of data specification. Each horizontal line represents one sentence of speech. Selected quotes are marked by number.}
\label{fig:quotes}
\end{figure*}

In many cases, groups ultimately adopted the teacher-recommended labels, indicating an openness to identify with practitioner values of supporting student autonomy and avoiding punitive administrative repercussions. However, teachers occasionally received extensive push-back from technical stakeholders. Such back-and-forth patterns between teachers and machine learning engineers are illustrated in Figure~\ref{fig:quotes}. In these cases, teachers persisted in their advocacy until groups understood the intent and gravitas behind their concerns. In one session working with the engagement classification scenario, the teacher and engineer engaged in an extended heated exchange regarding the ethics of classifying students with emotion-based labels. Teacher T3 explained the racialized underpinnings of assuming the emotional states of students in classroom practice:

\begin{displayquote}
    T3 (Student Engagement Image Classification): ``I would personally feel like that's something I can decide based on myself and my rapport with the students. If a student was frustrated or confused, to use those labels, I would be concerned about stereotyping. It’s a really big problem in education, how black students versus white students, how their behaviors read to a lot of white teachers as different, even though it can just be their specific cultural background.'' (13)
\end{displayquote}

While advocating against using labels that make assumptions about the emotional states of students, teacher T3 alludes to two other sensitive themes in the education domain: the historical context of racially biased perceptions of student behavior, as well as ML's infringement on the teacher’s role of human judgment in the classroom. The exchange emphasized the weight of these critical tensions in education and the prominence of racial and cultural considerations in the domain. A similarly heated discussion in another session involved the collection of teaching quality evaluation data for predicting student drop-out risk. Teacher T5 argued that the use of teacher evaluation data by administrators would impact teacher unions and staffing policies, further complicating an existing district struggle with protections for teachers. By contextualizing the social and political constructs connected to technically objective variables, teacher advocacy enabled groups to collectively situate data decisions in a complex ecosystem and approach designs with respect and sensitivity toward the worldviews encapsulated.

Across many sessions, advocacy operated through the sharing of personal lived experiences. While designing dataset composition in the automated essay grading scenario, teacher T9 argued against relying on quantified rule-based grammatical features to evaluate student writing:

\begin{displayquote}
    T9 (Automated Essay Grading): ``When I taught English on the west side of Chicago, ninety-nine percent of my students were African American. I understood what they were saying, but it was not written sort of like traditional academic American English. You don't want to penalize the student for the culture that they live in, and the language that they speak.'' (14)
\end{displayquote}

Teacher T9 reflected on their personal struggle to both respect individual student backgrounds and prepare students for future strict evaluative environments, and admitted that they still feel uncertain about the balance. The problem-solving nature of the co-design sessions invited sensitive practitioner stories involving difficulties faced in the classroom. Despite the relative vulnerability required of teachers and students compared to other roles, they met the task and eagerly advocated for the complex realities in the domain. 

\subsubsection{Ambiguity}
During time-limited co-design sessions, participants navigated the balance between big-picture discussions and specifying design details. Engaging in high-level discussions required stakeholders to develop a sense of \textbf{comfort with ambiguity}, accepting data unknowns and unfinished design decisions. Though participants expressed uneasiness about ambiguity in the design task, engineer E5 noted how the process stands in contrast to designing with a false sense of certainty:

\begin{displayquote}
    E5 (Student Drop-out Risk Prediction): ``When we’re talking about designing a system everybody wants to pretend they know more than they think. When we talk about making a decision everybody feels like they already know the answer, like they should know the answer. In this kind of setup\ldots I feel that it’s okay for me to not know the answer\ldots and I rely on other roles.'' (15)
\end{displayquote}

The presence of diverse stakeholders facilitates a collective acceptance of ambiguity while choosing proactive big-picture data planning. Engineers in other sessions echoed their ultimate appreciation of the open-ended nature of design decisions, given the relative infrequency of higher-level conversations in typical machine learning practice.

\subsection{Shifting Roles, Identities, and Support Needs}
Our study surfaces role-based collaborative dynamics and persistent knowledge gaps and boundaries that \textit{complicate} contribution in multi-stakeholder settings. Although participants engaged productively in the co-design process, we observed groups making assumptions, building on misconceptions, and getting stumped by shared unknowns. Stakeholders struggled with role-based identities and contributions. We identify challenges and support needs for engaging diverse stakeholders in collaborative data specification and summarize these in the final column of Table~\ref{tab:downstream_context}.

\subsubsection{Rigid responsibility boundaries}
While co-design sessions encouraged many boundary-spanning practices between engineers and domain experts, some role-based boundaries persisted and hindered collaboration. Several engineers maintained a bounded view of responsibilities and liability in data decisions. Especially for evaluations of ethical decisions, fairness for demographic subgroups, and consent practices, engineers were quick to \textbf{delegate to specialized entities}. Regarding the representation of subgroups in the composition stage of the protocol, engineer E1 explained:

\begin{displayquote}
    E1 (Student Engagement Image Classification): ``This is usually something that you shouldn’t be asking just anybody. I’d leave this question up to the ethics review panel professionals.'' (16)
\end{displayquote}

In an industry made efficient through role-based specialization, engineering \textbf{responsibilities may be narrowly defined}. Ethical standards are often handled by designated professionals in teams separate from engineering processes. Beyond preferring a separation of concerns, engineers indicated being accustomed to a standard industry practice removed from dataset design choices. In the automated essay grading scenario, engineer E10 explained their unease in the data collection stage of the protocol, while deciding between several ideas for labeling schemes:

\begin{displayquote}
    E10 (Automated Essay Grading): ``We don’t usually have that many options. You don’t have a choice, we really work with what we have. I don’t design how people give their data.'' (17)
\end{displayquote}

While engineers engaged in knowledge-sharing and translating practices, some did so to varying degrees of effectiveness. Despite an engineering effort to break down technical barriers, some teachers continued to feel intimidated by technology blindness. Others further struggled to contribute when additionally perceiving a misalignment between the student age group they had experience teaching and the target student age group for the machine learning application. Teacher T4 explained:

\begin{displayquote}
    T4 (Resume-base Career Recommendation): ``I didn’t think I had as much to add from an education perspective because my background is with a lot younger students.'' (18)
\end{displayquote}

Despite differences in the age group or subject matter in which teachers are more experienced, teachers nonetheless contribute domain expertise. By underselling their understanding of the context from working in the education system as an educator more generally, some teachers viewed their own stakeholder role as a direct end-user rather than co-designer. In several sessions, teachers fell more silent to self-imposed boundaries and misconceptions of qualifications for participation.

\subsubsection{Persistent knowledge gaps}
Groups face knowledge gaps regarding the identification of data owners. Despite the collection of educational data from academic records, classroom observations, and student work, teachers were unsure of data ownership regulations. Referring to collecting student essay data, teacher T8 explained:

\begin{displayquote}
    T8 (Automated Essay Grading): ``They probably live in the teachers’ district or school Google Drive, or whatever they’re using\ldots and so I’m not really sure who owns the rights to those.'' (19)
\end{displayquote}

Data ownership is further complicated by the storage of data through ed-tech systems with opaque data privacy terms negotiated with school administration. Even for data assumed to be maintained directly by school administration, teachers could not identify ownership and access processes. With regard to obtaining student academic records, multiple teachers expressed the need for school administrators to be present. Circumstances in education systems introduce variance in roles and responsibilities across districts, complicating the task of bringing stakeholder voices to the table. For domain experts, this knowledge gap may point to an unknown stakeholder and data owner in school administration who should be engaged in designing data specifications.
Engineers similarly make assumptions about data availability. Regarding a variety of student academic and financial records, engineer E6 assumed:

\begin{displayquote}
    E6 (Student Drop-out Risk Prediction): ``A lot of this data can be directly collected through the College Application, right? Sounds like a safe bet to say the university has all of them, I mean they keep a record of everything. It’s probably already marked in their system.'' (20)
\end{displayquote}

For engineers, these assumptions may be reflective of their training, in which datasets are given rather than constructed. Despite unknowns regarding data use agreements, shape and composition of data, and the right data owner to contact, engineers maintained confidence that data exists.

Despite involving stakeholders advocating for the legality of data privacy issues, knowledge gaps persist in data security and access. Engineer E1 used an assumption of security to justify haphazard consent practices:

\begin{displayquote}
    E1: ``For educational applications where there's going to be very little chance of any kind of deleterious impact or data leak, … you can ask the children to check a box saying my parents approve.'' (21)
\end{displayquote}

Both technical and non-technical participants expressed relaxed attitudes toward data security, citing a lack of incentive for malicious data breaches. Teacher T2 explained, \textit{``I don’t see why anybody would even want to hack into something that the schools were using.''}

\subsubsection{Shifting stakeholder identities}

While the teacher, engineer, and legal professional held clear participatory roles with established expectations for contribution, the collaborative identities of the student and UX professional roles were less defined. The distinct participation patterns across participants is illustrated in Figure~\ref{fig:quotes}, indicating the consistent interactions between teachers and engineers in contrast to the variance in the contributions of students and UX professionals. Student S4 explained, \textit{``consistently occupying the student perspective\ldots is difficult because the students are not going to be involved in every stage.''} In several sessions, students expressed discomfort being the youngest in a group of professionals and lacking a defined structure of contribution. As a result, they often removed themselves from their primary role of end-user and contributed to the collective task in a general co-designer capacity. Students eagerly participated in technical co-design, offering data collection and modeling ideas unrelated to the student perspective. For example, while the group discussed potential features to include in the dataset composition stage of the protocol for the student drop-out prediction scenario, student S5 explained their ideas for unconventional survey methods:

\begin{displayquote}
    S5 (Student Drop-out Risk Prediction): ``it might be interesting to have students survey each other almost, and\ldots this feels kind of creepy, but assigning one person in each of your classes and then say `does this person seem okay or do they seem like they’re gonna fail out of school', I think that that could be interesting.”\\
E5 (Student Drop-out Risk Prediction): ``I could see the potential bullying material for that.'' (22)
\end{displayquote}

In an effort to contribute to the design task, the student had \textbf{lost empathy with data subjects}, requiring the engineer to point out the potential downstream harms. Several students in the student drop-out prediction scenario additionally advocated for collecting mental health and other sensitive information from students. Across all sessions, students often referred to data subjects as “they” and separate from themselves. 

We find that, across all sessions, stakeholders lean on prior roles and experiences, often demonstrating multiple competencies, and shifting between these identities throughout the collaborative co-design process. Several legal professionals additionally had technical experience such that they could contribute ML best practices and participate in translation practices. Several designers and engineers had prior experience in various positions in the education sector such that they could contribute ideas motivated by domain expertise to contextualize data. Every stakeholder either remembers the experience of having been a student or is closely connected with a student through their social relationships, such that they could speak on behalf of student interests. Meanwhile, students struggled to always contribute through the role of a data subject and end-user. They instead often opted to participate through the role of a co-designer of data specifications.

UX professionals similarly lacked definition in their participatory roles. In the debriefing of the workshop, engineer E7 reflected, \textit{``this problem didn’t have too much to do with the UX perspective, so the collaborators have unequal representation in the discussion.''} By organizing co-design workshops that engage diverse stakeholders, the setting of the research study performed some of the traditional roles of designers, confusing the group's understanding of their expected contribution. Across sessions, several designers additionally maintained a role-based separation of concerns, limiting their contribution to technical topics. Removed from technical contribution and lacking the personal domain experiences of teachers and students, designers often faded into the background.

\section{Discussion}
Our findings demonstrate the vital role that multi-stakeholder collaborations play in the design of dataset specifications. Through conducting workshops anchored in the co-design of ML datasets in the high-stakes field of education, we highlight practitioner efforts to contextualize domain and procedural knowledge, establish common ground, and mitigate downstream harms. Participants engaged in a generative process of negotiating data requirements and quality in each stage of the data pipeline, placing due emphasis on the proactive design of human-centered systems. We emphasize the value of engaging domain experts and discuss the challenges facing the scalable implementation of the collaborative processes explored in this study. We characterize our contributions in terms of implications for the work of data specification and support needs for future integration of multi-stakeholder collaborative processes in responsible data use in education.  

\subsection{Is a Seat at the Table Enough?}
Critical scholarship has explored the tension between developing machine learning systems for scalable production and involving end users in the design of these systems~\cite{sloane2020participation}. By establishing a structured co-design environment in which diverse stakeholders were given a seat at the table, many of our participants engaged in organic negotiations to overcome knowledge boundaries to establish common ground through collaborative strategies. We found that participants both made significant contributions to the specification of data requirements and faced challenges in the co-design process. In this section, we discuss the affordances and limitations of stakeholder involvement in our workshop sessions through two participation frameworks and the application context of the education domain.

Delgado et al. describe an analytical framework for the dimensions of participation, designed for practitioners to assess the extent to which a process for participation meaningfully empowers diverse stakeholders in the design of ML applications~\cite{delgado2021stakeholder}. Consultation, involvement, collaboration, and empowerment are scaled degrees of participation, assessed at five decision points addressing the motivation, stakes, attendance, form, and power distribution in participation processes. Sloane et al. caution against ``participation washing,'' in which the narrative of participation obscures power dynamics and the extractive nature of collaboration~\cite{sloane2020participation}. By critically framing participatory design practices as work, consultation, and justice forms, practitioners may assess the authenticity of collaborative processes. Here we reflect on the nature of participation in our co-design sessions. 

\subsubsection{Affordances of co-design}
Our findings effectively demonstrate the crucial role of collaboration in data specifications, contributing to prior literature by identifying and confirming a critical site of participation in the development of ML applications. Across multiple decision points in Delgado et al.'s framework~\cite{delgado2021stakeholder}, the participatory structure of our workshops improves upon current practices, which limit teachers to lower degrees of participation, such as engaging in design feedback to improve the user experience of AI systems~\cite{holstein2019designing, roll2016evolution}. In contrast, the co-design of data specification is a participatory structure with stakes that empower stakeholders to contribute to the scope and purpose of ML applications. Data is the backbone of ML models and specifying data requirements is a systematic way to influence system behavior and hold AI accountable to stakeholders. By engaging stakeholders in this high-leverage high-impact stage of the ML pipeline, the design of data attributes and evaluation of data quality can systematically amplify the impact of stakeholder voices. 

Teachers shared domain expertise impacting critical features in dataset design and model performance. Out-of-domain experts commonly expressed surprise while recognizing their own knowledge gaps and disrupted assumptions regarding contextual variables relevant to the education domain. When reflecting on their collaborative experiences, engineers across all sessions discussed the narrow technical focus of traditional ML development processes, admitting frequent misplaced efforts and overlooked practitioner priorities. In most sessions, engineers expressed appreciation for the value of collaboration in the early stages of projects and eagerness to incorporate the process into practice. Teachers similarly expressed enthusiasm about contributing to data work. Stakeholders agreed unanimously across sessions that early stakeholder participation in designing data specifications uncovers domain-relevant priorities and potential downstream harms.

Groups additionally realized the value of ``big picture'' conversations to anticipate future harms by incorporating a contextual understanding of how data choices affect end-users and their environments. Rather than converging on design decisions, groups engaged in a process of ideation and filtration that often resulted in the recognition of multiple possibilities for further exploration. In line with collaborative approaches emphasizing the importance of friction and disagreement~\cite{keshavarz2013design}, diverging perspectives encouraged participants to develop an appreciation for ambiguity. The construction of a jointly negotiated framing, in a participatory process in which diverse stakeholder expertise is valued equally, promotes a plurality of designs in the resulting specifications. In the process, participants admit unknowns and rely on the collective knowledge of those at the table.

\subsubsection{Unequal burdens}
To engage in effective co-design, participants traversed expertise boundaries by practicing collaborative strategies unique to their roles. We observed domain experts striving to establish common ground by sharing vulnerable personal experiences and advocating for practitioner needs in a complicated historical and socio-cultural context. Out-of-domain stakeholders frequently and confidently made assumptions about the education system. Teachers and students are left with the emotional burden of advocating for their classroom experiences and navigating technology-centric pushback. In contrast, a heavy communicative burden is placed on the role of technical experts as boundary spanners.
This finding extends the existing understanding of ML team collaboration where technical members lack domain contexts, in which parties assume a diverse but equal contribution~\cite{piorkowski2021ai,muller2019data,park2021facilitating}.
Because domain practitioners in education feel significant barriers to contributing to the design of highly technical applications, engineers must support the central role of translation in collaborative practice. In the most positive collaborative sessions, engineers exhibited a willingness to teach foundational concepts and processes, patiently explaining technical tradeoffs. By translating domain-specific requirements to data and modeling requirements, engineers were able to address the needs advocated by non-technical domain experts. However, despite the productive ends of these collaborative strategies, they place extensive capacity demands in unique ways on both technical and non-technical participants. Without structured support, multi-stakeholder collaboration is a high-lift endeavor. Across our sessions, groups recurrently fell back on engineer-led linear decision-making when any stakeholder lacked the skill, knowledge, or energy capacity to meet these collaborative requirements.

In Sloane et al.'s framing of the forms of participation, the co-design workshops in this study align most closely with ``participation as consultation,'' in which diverse stakeholders are engaged in various stages of episodic, short-term projects. Consultative forms of participation often take a one-size-fits-all approach, creating a single process and expecting the same form of contribution from all stakeholders. However, diverse stakeholders contribute differently and require different support. Collaborative processes can better engage stakeholder perspectives by designing participation to be context and stakeholder-specific, revisiting processes to ensure the appropriate information is given to and gathered from the appropriate stakeholders. 

\subsubsection{Unfilled seats}
Our co-design workshops represent a lower degree of participation along the dimension of stakeholder selection, as the included community members were chosen by the research team. According to Delgado et al.'s framework, a participation process that truly empowers stakeholders involves engaging community members designated by the community itself. While valuable and necessary, this standard is difficult to achieve in the education context.

The education domain involves complex dynamic systems affected by the attributes of institutions, practitioners, communities, and policy. While teachers contributed eagerly to the co-design of data specifications, participants across sessions grappled with the unknown perspectives of diverse domain-relevant stakeholders not represented in our workshops. Beyond students and teachers, groups named parents, district administrators, school counselors, instructional coaches, support staff and district personnel, community organizations, and policymakers as significant stakeholders in the design scenarios presented. Participants consistently demonstrated knowledge gaps regarding organizational structures in schools and school systems, struggling to identify data owners and match position titles to role-based responsibilities. While participants often referenced ``administration'' as an agentic entity and critical stakeholder, the specific practitioner role to call upon remains undefined. Further, teachers shared experiences in which they were required to perform the responsibilities of administration and support staff when those roles were unavailable. The complexity of education systems is exacerbated by constant organizational change, situational differences across individual institutions, and overlapping roles due to personnel turnover and resource strain. In the education domain, the non-trivial task of identifying the full set of appropriate stakeholders to bring to the table is a necessary precondition to multi-stakeholder collaboration.

\subsection{Supporting Collaborative Data Specification}
Prior work assessing engineering processes has noted the lack of defined practices for engaging domain experts and diverse stakeholders, as well as the implications for the fairness and utility of the resulting systems~\cite{subramonyam2021towards}. In response, we formulated our workshop protocol as a preliminary approach to multi-stakeholder co-design of data specifications. Our findings demonstrate that, in order for collaborative data specification to realize its potential of systematically supporting and amplifying diverse stakeholder voices, structural supports are required. We further contribute to the participatory design literature by identifying process needs that facilitate participation by establishing common ground and continuous collaboration practices.

\subsubsection{Information Scaffolds}
Prior to the co-design workshops, participants were only informed about the research motivations. The design scenario and background about data use in ML development were presented during the session. As a result, participation in our workshops took the form of facilitated group discussions initiated by researchers. While this research design enabled us to observe practitioners navigating knowledge gaps, the introduction of initial groundwork may enable a higher degree of participation. Intentionally designed information scaffolds can establish common ground, overcome technical knowledge gaps, and accelerate proactive contributions across participants.

While the lack of appropriate materials for educating domain experts on foundational ML knowledge is known~\cite{bogina2022educating}, materials for educating ML practitioners on domain context are equally scarce. Despite a common assumption in collaborative ML work that non-technical experts require scaffolding of technical information, the same assumptions, and requirements and rarely ascribed to technical experts. Pre-reading regarding the social and political context of the design scenario may seed a foundational understanding of domain needs, practices, and motivations. Further, participants reported feeling unsure about the quality of their contributions and uncertain about the expectations of their role. Establishing common ground and defining stakeholder roles prior to engaging in co-design may better prepare participants for richer collaborative discussions.

Information scaffolds may additionally define the collaborative context more effectively. Participants struggled with decision-making due to a lack of clarity regarding the constraints of the design scenario. While the open-ended scenario invited high-level negotiations of priorities and requirements, groups reflected on the potential value of bounding ideation with real-world conditions, factoring financial, labor, and time resources into the initial formulation of the task. Finally, the inclusion of initial groundwork may enhance the generative design process, giving participants time to ideate independently before joint discussion and decision-making. 

\subsubsection{Shared standards}
Across sessions, groups struggled the most with designing specifications for the evaluation of data quality. Participants found this task challenging due to the lack of shared language and metrics for data and system evaluation across disciplines. Non-technical stakeholders were unfamiliar with standard machine learning metrics, such as accuracy, prediction, and recall, and lacked context regarding their applied meanings for the given design scenario. Correspondingly, technical stakeholders were unable to translate practitioner calls for evaluations based on student learning outcomes into actionable data specifications. Despite recognizing the incompleteness of any existing metric, participants faced difficulties creating new ones that better serve contextual needs. We echo the call from prior work that fairness in ML requires the development of domain-specific metrics of quality~\cite{shi2020artificial}.

\subsubsection{Continuous iteration}
Applying Delgado et al.'s framework, participation in our workshops is a one-time collaboration needed to better align ML applications with stakeholder needs. While this motivation is representative of a high degree of collaborative participation, the design falls short of empowering stakeholders due to a lack of accountability processes. Without accountability for the quality of implementation of stakeholder contributions, participation in workshops can become performative, failing to actualize the recommendations of diverse stakeholders. According to Sloan et al.'s framework, the most meaningful form of stakeholder involvement (i.e., ``participation as justice'') requires long-term partnerships with diverse stakeholders, building trust through mutual benefit, reciprocity, equity, and tightly coupled relationships with frequent communication. To establish processes for cross-domain collaboration, prior work has highlighted the importance of design iteration with constant evaluation~\cite{subramonyam2021towards}.

Indeed, participants in every workshop session echoed this collaborative requirement, calling for stakeholder involvement at each step in the execution of the data specification. While domain practitioners appreciated the proactive data planning exercise, they expressed concerns about implementation fidelity and the potential for harmful assumptions to re-enter the development process in their absence. Involving multi-stakeholders in continuous collaboration requires the construction of a framework that defines and scaffolds participant roles in iterative data specification in downstream stages of the ML pipeline. Some groups suggested the collaborative creation of a governing set of utility and ethical standards to be used in interval quality evaluations as new data decisions and trade-offs emerge. Support for sustained participation may also involve the development of software platforms to engage stakeholders in the downstream processes of data cleaning and model evaluation. Industry applications of ML development may benefit from the creation of new roles, hiring teachers and boundary-spanners in permanent or semi-permanent positions. The emerging field of education data science may train individuals with expertise at the intersection of technology and education, who are able to translate across domains. Future work should explore these and other processes necessary to sustain iterative and long-term end-user and domain expert participation in data and machine learning development, in each stage of the data lifecycle and beyond. 

\subsection{Limitations}
We present our data specification workshop procedure as a proof of concept with the acknowledgment that our sessions are subject to several limitations. A 2-hour workshop represents an oversimplified data specification process in which time constraints affect the nature of participation. The narrative surrounding the involvement of diverse stakeholders focuses on the prevention of harm by engaging domain context and impacted communities (e.g.,~\cite{boyarskaya2020overcoming}). Such framing of priority, combined with time constraints and latent structural inequality between stakeholder roles, limit the contribution of teachers to advocacy around well-recognized challenges in the education domain. When knowledge gaps are large enough, collaborative time is dedicated to establishing baseline domain understanding, leaving the full potential of stakeholder contribution unexplored. We chose to conduct design workshops despite literature acknowledging their limitations in participatory design \cite{rosner2016out, harrington2019deconstructing} because they allow us to imagine the research methods that may be adapted to authentic contexts. The demonstration presented here cannot assess the efficiency, feasibility, or affordability of collaborative data specification procedures. However, we identify their promise in addressing downstream issues of model-centric development and invite future work to explore the integration of this practice in industry settings.

Furthermore, our sampling methods may have introduced selection bias, favoring stakeholders who felt fewer barriers to participation and displayed an eagerness to seek collaborative work. Our participants additionally comprised an incomplete representation of stakeholders' communities. Due to convenience and ethical regulations regarding participation in research, the student role was represented by undergraduate students, rather than younger students more accurately impacted by the largely K-12 design scenarios. We recruited ML engineers from a variety of research and industry backgrounds, and teachers specializing in various age groups and subject disciplines, and we do not account for the role of this heterogeneity in the participatory results.

Finally, the four design scenarios produced heterogeneity across sessions and introduced different challenges and design discussions due to the nature of the datasets involved (e.g., tabular, text, and image data). Some scenarios were more emotionally charged, while others were more challenging technically to stakeholders. For example, the undertone of surveillance in scenarios involving student images prompted richer discourse about racial biases, representation, and fairness than in scenarios involving text data. While tabular data was easier to conceptualize, stakeholders were less familiar with image and text processing for data cleaning procedures, and these sessions relied heavily on technical experts to explain processing.

\section{Conclusion}
The emerging fairness, accountability, transparency, and utility concerns surrounding the development of ML applications in education are rooted in the limitations of conventional ML engineering processes. Developing ethical and human-centered ML experiences for education scenarios requires the prioritization of high-quality data contextualized through early collaboration with teachers and students. By engaging diverse stakeholders in a series of co-design sessions, we observed meaningful contributions to dataset specification. Participants shared domain and technical expertise to contextualize data needs, advocate for stakeholder values, anticipate downstream implications, overcome knowledge boundaries, and establish common ground. However, despite the many affordances of our collaborative process, a seat at the table is not enough. Empowering stakeholder perspectives in ML dataset specification requires systematic support, including accountable processes for the continuous involvement of teachers and students in iteration and co-evaluation, shared contextual data quality standards, and information scaffolds for both technical and non-technical stakeholders to traverse expertise boundaries.

\begin{acks}
We are grateful to the reviewers and our study participants for their time and helpful feedback. This work was funded through a grant from the McCoy Family Center for Ethics in Society at Stanford University. Dakuo Wang was supported by IBM Research as a visiting researcher at Stanford's Human-Centered AI Institute. 
\end{acks}

\bibliographystyle{ACM-Reference-Format}
\bibliography{99_refs}


\begin{thebibliography}{100}


\ifx \showCODEN    \undefined \def \showCODEN     #1{\unskip}     \fi
\ifx \showDOI      \undefined \def \showDOI       #1{#1}\fi
\ifx \showISBNx    \undefined \def \showISBNx     #1{\unskip}     \fi
\ifx \showISBNxiii \undefined \def \showISBNxiii  #1{\unskip}     \fi
\ifx \showISSN     \undefined \def \showISSN      #1{\unskip}     \fi
\ifx \showLCCN     \undefined \def \showLCCN      #1{\unskip}     \fi
\ifx \shownote     \undefined \def \shownote      #1{#1}          \fi
\ifx \showarticletitle \undefined \def \showarticletitle #1{#1}   \fi
\ifx \showURL      \undefined \def \showURL       {\relax}        \fi
\providecommand\bibfield[2]{#2}
\providecommand\bibinfo[2]{#2}
\providecommand\natexlab[1]{#1}
\providecommand\showeprint[2][]{arXiv:#2}

\bibitem[Akkiraju et~al\mbox{.}(2020)]%
        {akkiraju2020characterizing}
\bibfield{author}{\bibinfo{person}{Rama Akkiraju}, \bibinfo{person}{Vibha
  Sinha}, \bibinfo{person}{Anbang Xu}, \bibinfo{person}{Jalal Mahmud},
  \bibinfo{person}{Pritam Gundecha}, \bibinfo{person}{Zhe Liu},
  \bibinfo{person}{Xiaotong Liu}, {and} \bibinfo{person}{John Schumacher}.}
  \bibinfo{year}{2020}\natexlab{}.
\newblock \showarticletitle{Characterizing machine learning processes: A
  maturity framework}. In \bibinfo{booktitle}{\emph{International Conference on
  Business Process Management}}. Springer, \bibinfo{pages}{17--31}.
\newblock


\bibitem[Amershi et~al\mbox{.}(2019)]%
        {amershi2019software}
\bibfield{author}{\bibinfo{person}{Saleema Amershi}, \bibinfo{person}{Andrew
  Begel}, \bibinfo{person}{Christian Bird}, \bibinfo{person}{Robert DeLine},
  \bibinfo{person}{Harald Gall}, \bibinfo{person}{Ece Kamar},
  \bibinfo{person}{Nachiappan Nagappan}, \bibinfo{person}{Besmira Nushi}, {and}
  \bibinfo{person}{Thomas Zimmermann}.} \bibinfo{year}{2019}\natexlab{}.
\newblock \showarticletitle{Software engineering for machine learning: A case
  study}. In \bibinfo{booktitle}{\emph{2019 IEEE/ACM 41st International
  Conference on Software Engineering: Software Engineering in Practice
  (ICSE-SEIP)}}. IEEE, \bibinfo{pages}{291--300}.
\newblock


\bibitem[Arnold et~al\mbox{.}(2019)]%
        {arnold2019factsheets}
\bibfield{author}{\bibinfo{person}{Matthew Arnold}, \bibinfo{person}{Rachel~KE
  Bellamy}, \bibinfo{person}{Michael Hind}, \bibinfo{person}{Stephanie Houde},
  \bibinfo{person}{Sameep Mehta}, \bibinfo{person}{Aleksandra Mojsilovi{\'c}},
  \bibinfo{person}{Ravi Nair}, \bibinfo{person}{K~Natesan Ramamurthy},
  \bibinfo{person}{Alexandra Olteanu}, \bibinfo{person}{David Piorkowski},
  {et~al\mbox{.}}} \bibinfo{year}{2019}\natexlab{}.
\newblock \showarticletitle{FactSheets: Increasing trust in AI services through
  supplier's declarations of conformity}.
\newblock \bibinfo{journal}{\emph{IBM Journal of Research and Development}}
  \bibinfo{volume}{63}, \bibinfo{number}{4/5} (\bibinfo{year}{2019}),
  \bibinfo{pages}{6--1}.
\newblock


\bibitem[Ashmore et~al\mbox{.}(2021)]%
        {ashmore2021assuring}
\bibfield{author}{\bibinfo{person}{Rob Ashmore}, \bibinfo{person}{Radu
  Calinescu}, {and} \bibinfo{person}{Colin Paterson}.}
  \bibinfo{year}{2021}\natexlab{}.
\newblock \showarticletitle{Assuring the machine learning lifecycle:
  Desiderata, methods, and challenges}.
\newblock \bibinfo{journal}{\emph{ACM Computing Surveys (CSUR)}}
  \bibinfo{volume}{54}, \bibinfo{number}{5} (\bibinfo{year}{2021}),
  \bibinfo{pages}{1--39}.
\newblock


\bibitem[Baker and Hawn(2022a)]%
        {baker2022algorithmic}
\bibfield{author}{\bibinfo{person}{Ryan~S Baker} {and} \bibinfo{person}{Aaron
  Hawn}.} \bibinfo{year}{2022}\natexlab{a}.
\newblock \showarticletitle{Algorithmic bias in education}.
\newblock \bibinfo{journal}{\emph{International Journal of Artificial
  Intelligence in Education}} \bibinfo{volume}{32}, \bibinfo{number}{4}
  (\bibinfo{year}{2022}), \bibinfo{pages}{1052--1092}.
\newblock


\bibitem[Baker and Hawn(2022b)]%
        {baker_algorithmic_2022}
\bibfield{author}{\bibinfo{person}{Ryan~S. Baker} {and} \bibinfo{person}{Aaron
  Hawn}.} \bibinfo{year}{2022}\natexlab{b}.
\newblock \showarticletitle{Algorithmic {Bias} in {Education}}.
\newblock \bibinfo{journal}{\emph{International Journal of Artificial
  Intelligence in Education}} \bibinfo{volume}{32}, \bibinfo{number}{4}
  (\bibinfo{date}{Dec.} \bibinfo{year}{2022}), \bibinfo{pages}{1052--1092}.
\newblock
\showISSN{1560-4306}
\urldef\tempurl%
\url{https://doi.org/10.1007/s40593-021-00285-9}
\showDOI{\tempurl}


\bibitem[Bender and Friedman(2018)]%
        {bender2018data}
\bibfield{author}{\bibinfo{person}{Emily~M Bender} {and} \bibinfo{person}{Batya
  Friedman}.} \bibinfo{year}{2018}\natexlab{}.
\newblock \showarticletitle{Data statements for natural language processing:
  Toward mitigating system bias and enabling better science}.
\newblock \bibinfo{journal}{\emph{Transactions of the Association for
  Computational Linguistics}}  \bibinfo{volume}{6} (\bibinfo{year}{2018}),
  \bibinfo{pages}{587--604}.
\newblock


\bibitem[Birhane et~al\mbox{.}(2021)]%
        {birhane2021values}
\bibfield{author}{\bibinfo{person}{Abeba Birhane}, \bibinfo{person}{Pratyusha
  Kalluri}, \bibinfo{person}{Dallas Card}, \bibinfo{person}{William Agnew},
  \bibinfo{person}{Ravit Dotan}, {and} \bibinfo{person}{Michelle Bao}.}
  \bibinfo{year}{2021}\natexlab{}.
\newblock \showarticletitle{The values encoded in machine learning research}.
\newblock \bibinfo{journal}{\emph{arXiv preprint arXiv:2106.15590}}
  (\bibinfo{year}{2021}).
\newblock


\bibitem[Birks et~al\mbox{.}(2008)]%
        {birks2008memoing}
\bibfield{author}{\bibinfo{person}{Melanie Birks}, \bibinfo{person}{Ysanne
  Chapman}, {and} \bibinfo{person}{Karen Francis}.}
  \bibinfo{year}{2008}\natexlab{}.
\newblock \showarticletitle{Memoing in qualitative research: Probing data and
  processes}.
\newblock \bibinfo{journal}{\emph{Journal of research in nursing}}
  \bibinfo{volume}{13}, \bibinfo{number}{1} (\bibinfo{year}{2008}),
  \bibinfo{pages}{68--75}.
\newblock


\bibitem[Bogina et~al\mbox{.}(2022)]%
        {bogina2022educating}
\bibfield{author}{\bibinfo{person}{Veronika Bogina}, \bibinfo{person}{Alan
  Hartman}, \bibinfo{person}{Tsvi Kuflik}, {and} \bibinfo{person}{Avital
  Shulner-Tal}.} \bibinfo{year}{2022}\natexlab{}.
\newblock \showarticletitle{Educating software and AI stakeholders about
  algorithmic fairness, accountability, transparency and ethics}.
\newblock \bibinfo{journal}{\emph{International Journal of Artificial
  Intelligence in Education}} \bibinfo{volume}{32}, \bibinfo{number}{3}
  (\bibinfo{year}{2022}), \bibinfo{pages}{808--833}.
\newblock


\bibitem[Boyarskaya et~al\mbox{.}(2020)]%
        {boyarskaya2020overcoming}
\bibfield{author}{\bibinfo{person}{Margarita Boyarskaya},
  \bibinfo{person}{Alexandra Olteanu}, {and} \bibinfo{person}{Kate Crawford}.}
  \bibinfo{year}{2020}\natexlab{}.
\newblock \showarticletitle{Overcoming failures of imagination in AI infused
  system development and deployment}.
\newblock \bibinfo{journal}{\emph{arXiv preprint arXiv:2011.13416}}
  (\bibinfo{year}{2020}).
\newblock


\bibitem[Boyd(2021)]%
        {boyd2021datasheets}
\bibfield{author}{\bibinfo{person}{Karen~L Boyd}.}
  \bibinfo{year}{2021}\natexlab{}.
\newblock \showarticletitle{Datasheets for Datasets help ML Engineers Notice
  and Understand Ethical Issues in Training Data}.
\newblock \bibinfo{journal}{\emph{Proceedings of the ACM on Human-Computer
  Interaction}} \bibinfo{volume}{5}, \bibinfo{number}{CSCW2}
  (\bibinfo{year}{2021}), \bibinfo{pages}{1--27}.
\newblock


\bibitem[Buddemeyer et~al\mbox{.}(2021)]%
        {buddemeyer2021words}
\bibfield{author}{\bibinfo{person}{Amanda Buddemeyer}, \bibinfo{person}{Erin
  Walker}, {and} \bibinfo{person}{Malihe Alikhani}.}
  \bibinfo{year}{2021}\natexlab{}.
\newblock \showarticletitle{Words of Wisdom: Representational Harms in Learning
  From AI Communication}.
\newblock \bibinfo{journal}{\emph{arXiv preprint arXiv:2111.08581}}
  (\bibinfo{year}{2021}).
\newblock


\bibitem[Challa et~al\mbox{.}(2020)]%
        {challa2020faulty}
\bibfield{author}{\bibinfo{person}{Harshitha Challa}, \bibinfo{person}{Nan
  Niu}, {and} \bibinfo{person}{Reese Johnson}.}
  \bibinfo{year}{2020}\natexlab{}.
\newblock \showarticletitle{Faulty requirements made valuable: On the role of
  data quality in deep learning}. In \bibinfo{booktitle}{\emph{2020 IEEE
  Seventh International Workshop on Artificial Intelligence for Requirements
  Engineering (AIRE)}}. IEEE, \bibinfo{pages}{61--69}.
\newblock


\bibitem[Chmielinski et~al\mbox{.}(2022)]%
        {chmielinski2022dataset}
\bibfield{author}{\bibinfo{person}{Kasia~S Chmielinski}, \bibinfo{person}{Sarah
  Newman}, \bibinfo{person}{Matt Taylor}, \bibinfo{person}{Josh Joseph},
  \bibinfo{person}{Kemi Thomas}, \bibinfo{person}{Jessica Yurkofsky}, {and}
  \bibinfo{person}{Yue~Chelsea Qiu}.} \bibinfo{year}{2022}\natexlab{}.
\newblock \showarticletitle{The dataset nutrition label (2nd Gen): Leveraging
  context to mitigate harms in artificial intelligence}.
\newblock \bibinfo{journal}{\emph{arXiv preprint arXiv:2201.03954}}
  (\bibinfo{year}{2022}).
\newblock


\bibitem[Delgado et~al\mbox{.}(2021)]%
        {delgado2021stakeholder}
\bibfield{author}{\bibinfo{person}{Fernando Delgado}, \bibinfo{person}{Stephen
  Yang}, \bibinfo{person}{Michael Madaio}, {and} \bibinfo{person}{Qian Yang}.}
  \bibinfo{year}{2021}\natexlab{}.
\newblock \showarticletitle{Stakeholder Participation in AI: Beyond" Add
  Diverse Stakeholders and Stir"}.
\newblock \bibinfo{journal}{\emph{arXiv preprint arXiv:2111.01122}}
  (\bibinfo{year}{2021}).
\newblock


\bibitem[Denzin et~al\mbox{.}(2023)]%
        {denzin2023sage}
\bibfield{author}{\bibinfo{person}{Norman~K Denzin}, \bibinfo{person}{Yvonna~S
  Lincoln}, \bibinfo{person}{Michael~D Giardina}, {and}
  \bibinfo{person}{Gaile~S Cannella}.} \bibinfo{year}{2023}\natexlab{}.
\newblock \bibinfo{booktitle}{\emph{The Sage handbook of qualitative
  research}}.
\newblock \bibinfo{publisher}{Sage publications}.
\newblock


\bibitem[D{\'\i}az et~al\mbox{.}(2022)]%
        {diaz2022crowdworksheets}
\bibfield{author}{\bibinfo{person}{Mark D{\'\i}az}, \bibinfo{person}{Ian
  Kivlichan}, \bibinfo{person}{Rachel Rosen}, \bibinfo{person}{Dylan Baker},
  \bibinfo{person}{Razvan Amironesei}, \bibinfo{person}{Vinodkumar
  Prabhakaran}, {and} \bibinfo{person}{Emily Denton}.}
  \bibinfo{year}{2022}\natexlab{}.
\newblock \showarticletitle{Crowdworksheets: Accounting for individual and
  collective identities underlying crowdsourced dataset annotation}. In
  \bibinfo{booktitle}{\emph{2022 ACM Conference on Fairness, Accountability,
  and Transparency}}. \bibinfo{pages}{2342--2351}.
\newblock


\bibitem[DiSalvo et~al\mbox{.}(2012)]%
        {disalvo2012participatory}
\bibfield{author}{\bibinfo{person}{Carl DiSalvo}, \bibinfo{person}{Andrew
  Clement}, {and} \bibinfo{person}{Volkmar Pipek}.}
  \bibinfo{year}{2012}\natexlab{}.
\newblock \showarticletitle{Participatory design for, with, and by
  communities}.
\newblock \bibinfo{journal}{\emph{International handbook of participatory
  design}} (\bibinfo{year}{2012}), \bibinfo{pages}{182--209}.
\newblock


\bibitem[D’Amour et~al\mbox{.}(2020)]%
        {d2020underspecification}
\bibfield{author}{\bibinfo{person}{Alexander D’Amour},
  \bibinfo{person}{Katherine Heller}, \bibinfo{person}{Dan Moldovan},
  \bibinfo{person}{Ben Adlam}, \bibinfo{person}{Babak Alipanahi},
  \bibinfo{person}{Alex Beutel}, \bibinfo{person}{Christina Chen},
  \bibinfo{person}{Jonathan Deaton}, \bibinfo{person}{Jacob Eisenstein},
  \bibinfo{person}{Matthew~D Hoffman}, {et~al\mbox{.}}}
  \bibinfo{year}{2020}\natexlab{}.
\newblock \showarticletitle{Underspecification presents challenges for
  credibility in modern machine learning}.
\newblock \bibinfo{journal}{\emph{Journal of Machine Learning Research}}
  (\bibinfo{year}{2020}).
\newblock


\bibitem[Estivill-Castro et~al\mbox{.}(2022)]%
        {estivill2022constructing}
\bibfield{author}{\bibinfo{person}{Vladimir Estivill-Castro},
  \bibinfo{person}{Eugene Gilmore}, {and} \bibinfo{person}{Ren{\'e} Hexel}.}
  \bibinfo{year}{2022}\natexlab{}.
\newblock \showarticletitle{Constructing Explainable Classifiers from the
  Start—Enabling Human-in-the Loop Machine Learning}.
\newblock \bibinfo{journal}{\emph{Information}} \bibinfo{volume}{13},
  \bibinfo{number}{10} (\bibinfo{year}{2022}), \bibinfo{pages}{464}.
\newblock


\bibitem[Feathers(2022b)]%
        {feathers-1}
\bibfield{author}{\bibinfo{person}{Todd Feathers}.} \bibinfo{year}{Jan 11,
  2022}\natexlab{b}.
\newblock \showarticletitle{This Private Equity Firm Is Amassing Companies That
  Collect Data on America’s Children}.
\newblock \bibinfo{journal}{\emph{The Markup}} (\bibinfo{year}{Jan 11, 2022}).
\newblock
\newblock
\shownote{https://themarkup.org/machine-learning/2022/01/11/this-private-equity-firm-is-amassing-companies-that-collect-data-on-americas-children}.


\bibitem[Feathers(2022a)]%
        {feathers-2}
\bibfield{author}{\bibinfo{person}{Todd Feathers}.} \bibinfo{year}{Jan 13,
  2022}\natexlab{a}.
\newblock \showarticletitle{College Prep Software Naviance Is Selling
  Advertising Access to Millions of Students}.
\newblock \bibinfo{journal}{\emph{The Markup}} (\bibinfo{year}{Jan 13, 2022}).
\newblock
\newblock
\shownote{https://themarkup.org/machine-learning/2022/01/13/college-prep-software-naviance-is-selling-advertising-access-to-millions-of-students}.


\bibitem[Gardner et~al\mbox{.}(2019)]%
        {gardner2019evaluating}
\bibfield{author}{\bibinfo{person}{Josh Gardner}, \bibinfo{person}{Christopher
  Brooks}, {and} \bibinfo{person}{Ryan Baker}.}
  \bibinfo{year}{2019}\natexlab{}.
\newblock \showarticletitle{Evaluating the fairness of predictive student
  models through slicing analysis}. In \bibinfo{booktitle}{\emph{Proceedings of
  the 9th international conference on learning analytics \& knowledge}}.
  \bibinfo{pages}{225--234}.
\newblock


\bibitem[Gebru et~al\mbox{.}(2021)]%
        {gebru2021datasheets}
\bibfield{author}{\bibinfo{person}{Timnit Gebru}, \bibinfo{person}{Jamie
  Morgenstern}, \bibinfo{person}{Briana Vecchione},
  \bibinfo{person}{Jennifer~Wortman Vaughan}, \bibinfo{person}{Hanna Wallach},
  \bibinfo{person}{Hal~Daum{\'e} Iii}, {and} \bibinfo{person}{Kate Crawford}.}
  \bibinfo{year}{2021}\natexlab{}.
\newblock \showarticletitle{Datasheets for datasets}.
\newblock \bibinfo{journal}{\emph{Commun. ACM}} \bibinfo{volume}{64},
  \bibinfo{number}{12} (\bibinfo{year}{2021}), \bibinfo{pages}{86--92}.
\newblock


\bibitem[GmbH(2023)]%
        {atlas.ti}
\bibfield{author}{\bibinfo{person}{ATLAS.ti Scientific Software~Development
  GmbH}.} \bibinfo{year}{2023}\natexlab{}.
\newblock \bibinfo{title}{ATLAS.ti: The Qualitative Data Analysis \& Research
  Software}.
\newblock
\newblock
\urldef\tempurl%
\url{https://atlasti.com/}
\showURL{%
\tempurl}


\bibitem[Google(2019)]%
        {pair2019}
\bibfield{author}{\bibinfo{person}{Google}.} \bibinfo{year}{2019}\natexlab{}.
\newblock \bibinfo{title}{People + AI Guidebook}.
\newblock
\newblock
\urldef\tempurl%
\url{https://pair.withgoogle.com/}
\showURL{%
\tempurl}


\bibitem[Guo(2013)]%
        {guo2013data}
\bibfield{author}{\bibinfo{person}{Philip Guo}.}
  \bibinfo{year}{2013}\natexlab{}.
\newblock \showarticletitle{Data science workflow: Overview and challenges}.
\newblock \bibinfo{journal}{\emph{Commun. ACM}} (\bibinfo{year}{2013}).
\newblock


\bibitem[Harrington et~al\mbox{.}(2019)]%
        {harrington2019deconstructing}
\bibfield{author}{\bibinfo{person}{Christina Harrington},
  \bibinfo{person}{Sheena Erete}, {and} \bibinfo{person}{Anne~Marie Piper}.}
  \bibinfo{year}{2019}\natexlab{}.
\newblock \showarticletitle{Deconstructing community-based collaborative
  design: Towards more equitable participatory design engagements}.
\newblock \bibinfo{journal}{\emph{Proceedings of the ACM on Human-Computer
  Interaction}} \bibinfo{volume}{3}, \bibinfo{number}{CSCW}
  (\bibinfo{year}{2019}), \bibinfo{pages}{1--25}.
\newblock


\bibitem[Heger et~al\mbox{.}(2022)]%
        {heger2022understanding}
\bibfield{author}{\bibinfo{person}{Amy~K Heger}, \bibinfo{person}{Liz~B
  Marquis}, \bibinfo{person}{Mihaela Vorvoreanu}, \bibinfo{person}{Hanna
  Wallach}, {and} \bibinfo{person}{Jennifer Wortman~Vaughan}.}
  \bibinfo{year}{2022}\natexlab{}.
\newblock \showarticletitle{Understanding Machine Learning Practitioners' Data
  Documentation Perceptions, Needs, Challenges, and Desiderata}.
\newblock \bibinfo{journal}{\emph{Proceedings of the ACM on Human-Computer
  Interaction}} \bibinfo{volume}{6}, \bibinfo{number}{CSCW2}
  (\bibinfo{year}{2022}), \bibinfo{pages}{1--29}.
\newblock


\bibitem[Holmes et~al\mbox{.}(2022)]%
        {holmes2022ethics}
\bibfield{author}{\bibinfo{person}{Wayne Holmes}, \bibinfo{person}{Kaska
  Porayska-Pomsta}, \bibinfo{person}{Ken Holstein}, \bibinfo{person}{Emma
  Sutherland}, \bibinfo{person}{Toby Baker}, \bibinfo{person}{Simon~Buckingham
  Shum}, \bibinfo{person}{Olga~C Santos}, \bibinfo{person}{Mercedes~T Rodrigo},
  \bibinfo{person}{Mutlu Cukurova}, \bibinfo{person}{Ig~Ibert Bittencourt},
  {et~al\mbox{.}}} \bibinfo{year}{2022}\natexlab{}.
\newblock \showarticletitle{Ethics of AI in education: Towards a community-wide
  framework}.
\newblock \bibinfo{journal}{\emph{International Journal of Artificial
  Intelligence in Education}} \bibinfo{volume}{32}, \bibinfo{number}{3}
  (\bibinfo{year}{2022}), \bibinfo{pages}{504--526}.
\newblock


\bibitem[Holstein et~al\mbox{.}(2019a)]%
        {holstein2019co}
\bibfield{author}{\bibinfo{person}{Kenneth Holstein}, \bibinfo{person}{Bruce~M
  McLaren}, {and} \bibinfo{person}{Vincent Aleven}.}
  \bibinfo{year}{2019}\natexlab{a}.
\newblock \showarticletitle{Co-designing a real-time classroom orchestration
  tool to support teacher--AI complementarity}.
\newblock \bibinfo{journal}{\emph{Journal of Learning Analytics}}
  \bibinfo{volume}{6}, \bibinfo{number}{2} (\bibinfo{year}{2019}).
\newblock


\bibitem[Holstein et~al\mbox{.}(2019b)]%
        {holstein2019designing}
\bibfield{author}{\bibinfo{person}{Kenneth Holstein}, \bibinfo{person}{Bruce~M
  McLaren}, {and} \bibinfo{person}{Vincent Aleven}.}
  \bibinfo{year}{2019}\natexlab{b}.
\newblock \showarticletitle{Designing for complementarity: Teacher and student
  needs for orchestration support in AI-enhanced classrooms}. In
  \bibinfo{booktitle}{\emph{International conference on artificial intelligence
  in education}}. Springer, \bibinfo{pages}{157--171}.
\newblock


\bibitem[Holstein et~al\mbox{.}(2019c)]%
        {holstein2019improving}
\bibfield{author}{\bibinfo{person}{Kenneth Holstein}, \bibinfo{person}{Jennifer
  Wortman~Vaughan}, \bibinfo{person}{Hal Daum{\'e}~III}, \bibinfo{person}{Miro
  Dudik}, {and} \bibinfo{person}{Hanna Wallach}.}
  \bibinfo{year}{2019}\natexlab{c}.
\newblock \showarticletitle{Improving fairness in machine learning systems:
  What do industry practitioners need?}. In
  \bibinfo{booktitle}{\emph{Proceedings of the 2019 CHI conference on human
  factors in computing systems}}. \bibinfo{pages}{1--16}.
\newblock


\bibitem[Hou and Wang(2017)]%
        {hou2017hacking}
\bibfield{author}{\bibinfo{person}{Youyang Hou} {and} \bibinfo{person}{Dakuo
  Wang}.} \bibinfo{year}{2017}\natexlab{}.
\newblock \showarticletitle{Hacking with NPOs: collaborative analytics and
  broker roles in civic data hackathons}.
\newblock \bibinfo{journal}{\emph{Proceedings of the ACM on Human-Computer
  Interaction}} \bibinfo{volume}{1}, \bibinfo{number}{CSCW}
  (\bibinfo{year}{2017}), \bibinfo{pages}{1--16}.
\newblock


\bibitem[Hullman et~al\mbox{.}(2022)]%
        {hullman2022worst}
\bibfield{author}{\bibinfo{person}{Jessica Hullman}, \bibinfo{person}{Sayash
  Kapoor}, \bibinfo{person}{Priyanka Nanayakkara}, \bibinfo{person}{Andrew
  Gelman}, {and} \bibinfo{person}{Arvind Narayanan}.}
  \bibinfo{year}{2022}\natexlab{}.
\newblock \showarticletitle{The worst of both worlds: A comparative analysis of
  errors in learning from data in psychology and machine learning}.
\newblock \bibinfo{journal}{\emph{arXiv preprint arXiv:2203.06498}}
  (\bibinfo{year}{2022}).
\newblock


\bibitem[Hutchinson et~al\mbox{.}(2021)]%
        {10.1145/3442188.3445918}
\bibfield{author}{\bibinfo{person}{Ben Hutchinson}, \bibinfo{person}{Andrew
  Smart}, \bibinfo{person}{Alex Hanna}, \bibinfo{person}{Emily Denton},
  \bibinfo{person}{Christina Greer}, \bibinfo{person}{Oddur Kjartansson},
  \bibinfo{person}{Parker Barnes}, {and} \bibinfo{person}{Margaret Mitchell}.}
  \bibinfo{year}{2021}\natexlab{}.
\newblock \showarticletitle{Towards Accountability for Machine Learning
  Datasets: Practices from Software Engineering and Infrastructure}. In
  \bibinfo{booktitle}{\emph{Proceedings of the 2021 ACM Conference on Fairness,
  Accountability, and Transparency}} (Virtual Event, Canada)
  \emph{(\bibinfo{series}{FAccT '21})}. \bibinfo{publisher}{Association for
  Computing Machinery}, \bibinfo{address}{New York, NY, USA},
  \bibinfo{pages}{560–575}.
\newblock
\showISBNx{9781450383097}
\urldef\tempurl%
\url{https://doi.org/10.1145/3442188.3445918}
\showDOI{\tempurl}


\bibitem[Kerner(2020)]%
        {kerner2020too}
\bibfield{author}{\bibinfo{person}{Hannah Kerner}.}
  \bibinfo{year}{2020}\natexlab{}.
\newblock \showarticletitle{Too many AI researchers think real-world problems
  are not relevant}.
\newblock \bibinfo{journal}{\emph{Opinion. MIT Technology Review}}
  (\bibinfo{year}{2020}).
\newblock


\bibitem[Keshavarz and Maze(2013)]%
        {keshavarz2013design}
\bibfield{author}{\bibinfo{person}{Mahmoud Keshavarz} {and}
  \bibinfo{person}{Ramia Maze}.} \bibinfo{year}{2013}\natexlab{}.
\newblock \showarticletitle{Design and dissensus: framing and staging
  participation in design research}.
\newblock \bibinfo{journal}{\emph{Design Philosophy Papers}}
  \bibinfo{volume}{11}, \bibinfo{number}{1} (\bibinfo{year}{2013}),
  \bibinfo{pages}{7--29}.
\newblock


\bibitem[Kizilcec and Lee(2022)]%
        {aied_chapter}
\bibfield{author}{\bibinfo{person}{Ren\'e~F. Kizilcec} {and}
  \bibinfo{person}{Hansol Lee}.} \bibinfo{year}{2022}\natexlab{}.
\newblock \showarticletitle{Algorithmic Fairness in Education}.
\newblock In \bibinfo{booktitle}{\emph{The Ethics of Artificial Intelligence in
  Education}}, \bibfield{editor}{\bibinfo{person}{W.~Holmes \&~K.
  Porayska-Pomsta}} (Ed.). \bibinfo{publisher}{Routledge}, Chapter~7.
\newblock


\bibitem[Koesten et~al\mbox{.}(2021)]%
        {koesten2021talking}
\bibfield{author}{\bibinfo{person}{Laura Koesten}, \bibinfo{person}{Kathleen
  Gregory}, \bibinfo{person}{Paul Groth}, {and} \bibinfo{person}{Elena
  Simperl}.} \bibinfo{year}{2021}\natexlab{}.
\newblock \showarticletitle{Talking datasets--understanding data sensemaking
  behaviours}.
\newblock \bibinfo{journal}{\emph{International journal of human-computer
  studies}}  \bibinfo{volume}{146} (\bibinfo{year}{2021}),
  \bibinfo{pages}{102562}.
\newblock


\bibitem[Koopmans(2020)]%
        {koopmans2020education}
\bibfield{author}{\bibinfo{person}{Matthijs Koopmans}.}
  \bibinfo{year}{2020}\natexlab{}.
\newblock \showarticletitle{Education is a complex dynamical system: Challenges
  for research}.
\newblock \bibinfo{journal}{\emph{The Journal of Experimental Education}}
  \bibinfo{volume}{88}, \bibinfo{number}{3} (\bibinfo{year}{2020}),
  \bibinfo{pages}{358--374}.
\newblock


\bibitem[Kross and Guo(2021)]%
        {kross2021orienting}
\bibfield{author}{\bibinfo{person}{Sean Kross} {and} \bibinfo{person}{Philip
  Guo}.} \bibinfo{year}{2021}\natexlab{}.
\newblock \showarticletitle{Orienting, framing, bridging, magic, and
  counseling: How data scientists navigate the outer loop of client
  collaborations in industry and academia}.
\newblock \bibinfo{journal}{\emph{Proceedings of the ACM on Human-Computer
  Interaction}} \bibinfo{volume}{5}, \bibinfo{number}{CSCW2}
  (\bibinfo{year}{2021}), \bibinfo{pages}{1--28}.
\newblock


\bibitem[Latonero et~al\mbox{.}(2017)]%
        {latonero2017tech}
\bibfield{author}{\bibinfo{person}{M Latonero}, \bibinfo{person}{M Kleinman},
  {and} \bibinfo{person}{K Hiatt}.} \bibinfo{year}{2017}\natexlab{}.
\newblock \showarticletitle{Tech Folk:‘Move Fast and Break Things’
  Doesn’t Work When Lives Are at Stake}.
\newblock \bibinfo{journal}{\emph{The Guardian, February}}
  (\bibinfo{year}{2017}).
\newblock


\bibitem[Leslie et~al\mbox{.}(2022)]%
        {leslie2022data}
\bibfield{author}{\bibinfo{person}{David Leslie}, \bibinfo{person}{Michael
  Katell}, \bibinfo{person}{Mhairi Aitken}, \bibinfo{person}{Jatinder Singh},
  \bibinfo{person}{Morgan Briggs}, \bibinfo{person}{Rosamund Powell},
  \bibinfo{person}{Cami Rinc{\'o}n}, \bibinfo{person}{Antonella Perini},
  \bibinfo{person}{Smera Jayadeva}, {and} \bibinfo{person}{Christopher Burr}.}
  \bibinfo{year}{2022}\natexlab{}.
\newblock \showarticletitle{Data Justice in Practice: A Guide for Developers}.
\newblock \bibinfo{journal}{\emph{arXiv preprint arXiv:2205.01037}}
  (\bibinfo{year}{2022}).
\newblock


\bibitem[Li et~al\mbox{.}(2022)]%
        {li2022disparities}
\bibfield{author}{\bibinfo{person}{Warren Li}, \bibinfo{person}{Kaiwen Sun},
  \bibinfo{person}{Florian Schaub}, {and} \bibinfo{person}{Christopher
  Brooks}.} \bibinfo{year}{2022}\natexlab{}.
\newblock \showarticletitle{Disparities in students’ propensity to consent to
  learning analytics}.
\newblock \bibinfo{journal}{\emph{International Journal of Artificial
  Intelligence in Education}} \bibinfo{volume}{32}, \bibinfo{number}{3}
  (\bibinfo{year}{2022}), \bibinfo{pages}{564--608}.
\newblock


\bibitem[Liang et~al\mbox{.}(2022)]%
        {liang2022advances}
\bibfield{author}{\bibinfo{person}{Weixin Liang}, \bibinfo{person}{Girmaw~Abebe
  Tadesse}, \bibinfo{person}{Daniel Ho}, \bibinfo{person}{L Fei-Fei},
  \bibinfo{person}{Matei Zaharia}, \bibinfo{person}{Ce Zhang}, {and}
  \bibinfo{person}{James Zou}.} \bibinfo{year}{2022}\natexlab{}.
\newblock \showarticletitle{Advances, challenges and opportunities in creating
  data for trustworthy AI}.
\newblock \bibinfo{journal}{\emph{Nature Machine Intelligence}}
  \bibinfo{volume}{4}, \bibinfo{number}{8} (\bibinfo{year}{2022}),
  \bibinfo{pages}{669--677}.
\newblock


\bibitem[Mao et~al\mbox{.}(2019)]%
        {mao2019data}
\bibfield{author}{\bibinfo{person}{Yaoli Mao}, \bibinfo{person}{Dakuo Wang},
  \bibinfo{person}{Michael Muller}, \bibinfo{person}{Kush~R Varshney},
  \bibinfo{person}{Ioana Baldini}, \bibinfo{person}{Casey Dugan}, {and}
  \bibinfo{person}{Aleksandra Mojsilovi{\'c}}.}
  \bibinfo{year}{2019}\natexlab{}.
\newblock \showarticletitle{How data scientistswork together with domain
  experts in scientific collaborations: To find the right answer or to ask the
  right question?}
\newblock \bibinfo{journal}{\emph{Proceedings of the ACM on Human-Computer
  Interaction}} \bibinfo{volume}{3}, \bibinfo{number}{GROUP}
  (\bibinfo{year}{2019}), \bibinfo{pages}{1--23}.
\newblock


\bibitem[Marras et~al\mbox{.}(2022)]%
        {marras2022equality}
\bibfield{author}{\bibinfo{person}{Mirko Marras}, \bibinfo{person}{Ludovico
  Boratto}, \bibinfo{person}{Guilherme Ramos}, {and} \bibinfo{person}{Gianni
  Fenu}.} \bibinfo{year}{2022}\natexlab{}.
\newblock \showarticletitle{Equality of learning opportunity via individual
  fairness in personalized recommendations}.
\newblock \bibinfo{journal}{\emph{International Journal of Artificial
  Intelligence in Education}} \bibinfo{volume}{32}, \bibinfo{number}{3}
  (\bibinfo{year}{2022}), \bibinfo{pages}{636--684}.
\newblock


\bibitem[Michos et~al\mbox{.}(2020)]%
        {michos2020involving}
\bibfield{author}{\bibinfo{person}{Konstantinos Michos},
  \bibinfo{person}{Charles Lang}, \bibinfo{person}{Davinia Hern{\'a}ndez-Leo},
  {and} \bibinfo{person}{Detra Price-Dennis}.} \bibinfo{year}{2020}\natexlab{}.
\newblock \showarticletitle{Involving teachers in learning analytics design:
  Lessons learned from two case studies}. In
  \bibinfo{booktitle}{\emph{Proceedings of the Tenth international conference
  on learning analytics \& knowledge}}. \bibinfo{pages}{94--99}.
\newblock


\bibitem[Muller et~al\mbox{.}(2019)]%
        {muller2019data}
\bibfield{author}{\bibinfo{person}{Michael Muller}, \bibinfo{person}{Ingrid
  Lange}, \bibinfo{person}{Dakuo Wang}, \bibinfo{person}{David Piorkowski},
  \bibinfo{person}{Jason Tsay}, \bibinfo{person}{Q~Vera Liao},
  \bibinfo{person}{Casey Dugan}, {and} \bibinfo{person}{Thomas Erickson}.}
  \bibinfo{year}{2019}\natexlab{}.
\newblock \showarticletitle{How data science workers work with data: Discovery,
  capture, curation, design, creation}. In
  \bibinfo{booktitle}{\emph{Proceedings of the 2019 CHI conference on human
  factors in computing systems}}. \bibinfo{pages}{1--15}.
\newblock


\bibitem[Muller and Strohmayer(2022)]%
        {muller2022forgetting}
\bibfield{author}{\bibinfo{person}{Michael Muller} {and}
  \bibinfo{person}{Angelika Strohmayer}.} \bibinfo{year}{2022}\natexlab{}.
\newblock \showarticletitle{Forgetting Practices in the Data Sciences}. In
  \bibinfo{booktitle}{\emph{CHI Conference on Human Factors in Computing
  Systems}}. \bibinfo{pages}{1--19}.
\newblock


\bibitem[Munappy et~al\mbox{.}(2020)]%
        {munappy2020data}
\bibfield{author}{\bibinfo{person}{Aiswarya~Raj Munappy}, \bibinfo{person}{Jan
  Bosch}, {and} \bibinfo{person}{Helena~Homstr{\"o}m Olsson}.}
  \bibinfo{year}{2020}\natexlab{}.
\newblock \showarticletitle{Data pipeline management in practice: Challenges
  and opportunities}. In \bibinfo{booktitle}{\emph{International Conference on
  Product-Focused Software Process Improvement}}. Springer,
  \bibinfo{pages}{168--184}.
\newblock


\bibitem[Nagle et~al\mbox{.}(2017)]%
        {nagle2017only}
\bibfield{author}{\bibinfo{person}{Tadhg Nagle}, \bibinfo{person}{Thomas~C
  Redman}, {and} \bibinfo{person}{David Sammon}.}
  \bibinfo{year}{2017}\natexlab{}.
\newblock \showarticletitle{Only 3\% of companies’ data meets basic quality
  standards}.
\newblock \bibinfo{journal}{\emph{Harvard Business Review}}
  \bibinfo{volume}{95}, \bibinfo{number}{5} (\bibinfo{year}{2017}),
  \bibinfo{pages}{2--5}.
\newblock


\bibitem[Niemi et~al\mbox{.}(2023)]%
        {niemi2023ai}
\bibfield{author}{\bibinfo{person}{Hannele Niemi}, \bibinfo{person}{Roy~D Pea},
  {and} \bibinfo{person}{Yu Lu}.} \bibinfo{year}{2023}\natexlab{}.
\newblock \bibinfo{title}{AI in Learning: Designing the Future}.
\newblock
\newblock


\bibitem[Ocumpaugh et~al\mbox{.}(2014)]%
        {Ocumpaugh2014}
\bibfield{author}{\bibinfo{person}{Jaclyn Ocumpaugh}, \bibinfo{person}{Ryan
  Baker}, \bibinfo{person}{Sujith Gowda}, \bibinfo{person}{Neil Heffernan},
  {and} \bibinfo{person}{Cristina Heffernan}.} \bibinfo{year}{2014}\natexlab{}.
\newblock \showarticletitle{Population validity for educational data mining
  models: A case study in affect detection}.
\newblock \bibinfo{journal}{\emph{British Journal of Educational Technology}}
  \bibinfo{volume}{45} (\bibinfo{date}{05} \bibinfo{year}{2014}).
\newblock
\urldef\tempurl%
\url{https://doi.org/10.1111/bjet.12156}
\showDOI{\tempurl}


\bibitem[Park et~al\mbox{.}(2021)]%
        {park2021facilitating}
\bibfield{author}{\bibinfo{person}{Soya Park}, \bibinfo{person}{April~Yi Wang},
  \bibinfo{person}{Ban Kawas}, \bibinfo{person}{Q~Vera Liao},
  \bibinfo{person}{David Piorkowski}, {and} \bibinfo{person}{Marina
  Danilevsky}.} \bibinfo{year}{2021}\natexlab{}.
\newblock \showarticletitle{Facilitating knowledge sharing from domain experts
  to data scientists for building nlp models}. In
  \bibinfo{booktitle}{\emph{26th International Conference on Intelligent User
  Interfaces}}. \bibinfo{pages}{585--596}.
\newblock


\bibitem[Passi and Jackson(2018)]%
        {passi2018trust}
\bibfield{author}{\bibinfo{person}{Samir Passi} {and} \bibinfo{person}{Steven~J
  Jackson}.} \bibinfo{year}{2018}\natexlab{}.
\newblock \showarticletitle{Trust in data science: Collaboration, translation,
  and accountability in corporate data science projects}.
\newblock \bibinfo{journal}{\emph{Proceedings of the ACM on Human-Computer
  Interaction}} \bibinfo{volume}{2}, \bibinfo{number}{CSCW}
  (\bibinfo{year}{2018}), \bibinfo{pages}{1--28}.
\newblock


\bibitem[Passi and Sengers(2020)]%
        {passi2020making}
\bibfield{author}{\bibinfo{person}{Samir Passi} {and} \bibinfo{person}{Phoebe
  Sengers}.} \bibinfo{year}{2020}\natexlab{}.
\newblock \showarticletitle{Making data science systems work}.
\newblock \bibinfo{journal}{\emph{Big Data \& Society}} \bibinfo{volume}{7},
  \bibinfo{number}{2} (\bibinfo{year}{2020}),
  \bibinfo{pages}{2053951720939605}.
\newblock


\bibitem[Paullada et~al\mbox{.}(2021)]%
        {paullada2021data}
\bibfield{author}{\bibinfo{person}{Amandalynne Paullada},
  \bibinfo{person}{Inioluwa~Deborah Raji}, \bibinfo{person}{Emily~M Bender},
  \bibinfo{person}{Emily Denton}, {and} \bibinfo{person}{Alex Hanna}.}
  \bibinfo{year}{2021}\natexlab{}.
\newblock \showarticletitle{Data and its (dis) contents: A survey of dataset
  development and use in machine learning research}.
\newblock \bibinfo{journal}{\emph{Patterns}} \bibinfo{volume}{2},
  \bibinfo{number}{11} (\bibinfo{year}{2021}), \bibinfo{pages}{100336}.
\newblock


\bibitem[Pedro et~al\mbox{.}(2019)]%
        {pedro2019artificial}
\bibfield{author}{\bibinfo{person}{Francesc Pedro}, \bibinfo{person}{Miguel
  Subosa}, \bibinfo{person}{Axel Rivas}, {and} \bibinfo{person}{Paula
  Valverde}.} \bibinfo{year}{2019}\natexlab{}.
\newblock \showarticletitle{Artificial intelligence in education: Challenges
  and opportunities for sustainable development}.
\newblock  (\bibinfo{year}{2019}).
\newblock


\bibitem[Perrotta and Selwyn(2020)]%
        {perrotta2020deep}
\bibfield{author}{\bibinfo{person}{Carlo Perrotta} {and} \bibinfo{person}{Neil
  Selwyn}.} \bibinfo{year}{2020}\natexlab{}.
\newblock \showarticletitle{Deep learning goes to school: Toward a relational
  understanding of AI in education}.
\newblock \bibinfo{journal}{\emph{Learning, Media and Technology}}
  \bibinfo{volume}{45}, \bibinfo{number}{3} (\bibinfo{year}{2020}),
  \bibinfo{pages}{251--269}.
\newblock


\bibitem[Piet(2019)]%
        {aimeetsdesign}
\bibfield{author}{\bibinfo{person}{Nadia Piet}.}
  \bibinfo{year}{2019}\natexlab{}.
\newblock \bibinfo{title}{AI Meets Design}.
\newblock
\newblock
\urldef\tempurl%
\url{http://aimeets.design/}
\showURL{%
\tempurl}


\bibitem[Piorkowski et~al\mbox{.}(2021)]%
        {piorkowski2021ai}
\bibfield{author}{\bibinfo{person}{David Piorkowski}, \bibinfo{person}{Soya
  Park}, \bibinfo{person}{April~Yi Wang}, \bibinfo{person}{Dakuo Wang},
  \bibinfo{person}{Michael Muller}, {and} \bibinfo{person}{Felix Portnoy}.}
  \bibinfo{year}{2021}\natexlab{}.
\newblock \showarticletitle{How ai developers overcome communication challenges
  in a multidisciplinary team: A case study}.
\newblock \bibinfo{journal}{\emph{Proceedings of the ACM on Human-Computer
  Interaction}} \bibinfo{volume}{5}, \bibinfo{number}{CSCW1}
  (\bibinfo{year}{2021}), \bibinfo{pages}{1--25}.
\newblock


\bibitem[Polyzotis et~al\mbox{.}(2017)]%
        {polyzotis2017data}
\bibfield{author}{\bibinfo{person}{Neoklis Polyzotis}, \bibinfo{person}{Sudip
  Roy}, \bibinfo{person}{Steven~Euijong Whang}, {and} \bibinfo{person}{Martin
  Zinkevich}.} \bibinfo{year}{2017}\natexlab{}.
\newblock \showarticletitle{Data management challenges in production machine
  learning}. In \bibinfo{booktitle}{\emph{Proceedings of the 2017 ACM
  International Conference on Management of Data}}.
  \bibinfo{pages}{1723--1726}.
\newblock


\bibitem[Polyzotis et~al\mbox{.}(2018)]%
        {polyzotis2018data}
\bibfield{author}{\bibinfo{person}{Neoklis Polyzotis}, \bibinfo{person}{Sudip
  Roy}, \bibinfo{person}{Steven~Euijong Whang}, {and} \bibinfo{person}{Martin
  Zinkevich}.} \bibinfo{year}{2018}\natexlab{}.
\newblock \showarticletitle{Data lifecycle challenges in production machine
  learning: a survey}.
\newblock \bibinfo{journal}{\emph{ACM SIGMOD Record}} \bibinfo{volume}{47},
  \bibinfo{number}{2} (\bibinfo{year}{2018}), \bibinfo{pages}{17--28}.
\newblock


\bibitem[Potgieter(2020)]%
        {potgieter2020privacy}
\bibfield{author}{\bibinfo{person}{Isak Potgieter}.}
  \bibinfo{year}{2020}\natexlab{}.
\newblock \showarticletitle{Privacy concerns in educational data mining and
  learning analytics}.
\newblock \bibinfo{journal}{\emph{The International Review of Information
  Ethics}}  \bibinfo{volume}{28} (\bibinfo{year}{2020}).
\newblock


\bibitem[Pushkarna et~al\mbox{.}(2022)]%
        {pushkarna2022data}
\bibfield{author}{\bibinfo{person}{Mahima Pushkarna}, \bibinfo{person}{Andrew
  Zaldivar}, {and} \bibinfo{person}{Oddur Kjartansson}.}
  \bibinfo{year}{2022}\natexlab{}.
\newblock \showarticletitle{Data Cards: Purposeful and Transparent Dataset
  Documentation for Responsible AI}.
\newblock \bibinfo{journal}{\emph{arXiv preprint arXiv:2204.01075}}
  (\bibinfo{year}{2022}).
\newblock


\bibitem[Rakova et~al\mbox{.}(2021)]%
        {rakova2021responsible}
\bibfield{author}{\bibinfo{person}{Bogdana Rakova}, \bibinfo{person}{Jingying
  Yang}, \bibinfo{person}{Henriette Cramer}, {and} \bibinfo{person}{Rumman
  Chowdhury}.} \bibinfo{year}{2021}\natexlab{}.
\newblock \showarticletitle{Where responsible AI meets reality: Practitioner
  perspectives on enablers for shifting organizational practices}.
\newblock \bibinfo{journal}{\emph{Proceedings of the ACM on Human-Computer
  Interaction}} \bibinfo{volume}{5}, \bibinfo{number}{CSCW1}
  (\bibinfo{year}{2021}), \bibinfo{pages}{1--23}.
\newblock


\bibitem[Reich and Ito(2017)]%
        {reich2017good}
\bibfield{author}{\bibinfo{person}{Justin Reich} {and} \bibinfo{person}{Mizuko
  Ito}.} \bibinfo{year}{2017}\natexlab{}.
\newblock \showarticletitle{From good intentions to real outcomes: Equity by
  design in learning technologies}.
\newblock \bibinfo{journal}{\emph{Digital Media and Learning Research Hub}}
  (\bibinfo{year}{2017}).
\newblock


\bibitem[Richards et~al\mbox{.}(2021)]%
        {richards2021human}
\bibfield{author}{\bibinfo{person}{John~T Richards}, \bibinfo{person}{David
  Piorkowski}, \bibinfo{person}{Michael Hind}, \bibinfo{person}{Stephanie
  Houde}, \bibinfo{person}{Aleksandra Mojsilovic}, {and}
  \bibinfo{person}{Kush~R Varshney}.} \bibinfo{year}{2021}\natexlab{}.
\newblock \showarticletitle{A Human-Centered Methodology for Creating AI
  FactSheets.}
\newblock \bibinfo{journal}{\emph{IEEE Data Eng. Bull.}} \bibinfo{volume}{44},
  \bibinfo{number}{4} (\bibinfo{year}{2021}), \bibinfo{pages}{47--58}.
\newblock


\bibitem[Richardson et~al\mbox{.}(2019)]%
        {richardson2019dirty}
\bibfield{author}{\bibinfo{person}{Rashida Richardson},
  \bibinfo{person}{Jason~M Schultz}, {and} \bibinfo{person}{Kate Crawford}.}
  \bibinfo{year}{2019}\natexlab{}.
\newblock \showarticletitle{Dirty data, bad predictions: How civil rights
  violations impact police data, predictive policing systems, and justice}.
\newblock \bibinfo{journal}{\emph{NYUL Rev. Online}}  \bibinfo{volume}{94}
  (\bibinfo{year}{2019}), \bibinfo{pages}{15}.
\newblock


\bibitem[Roll and Wylie(2016)]%
        {roll2016evolution}
\bibfield{author}{\bibinfo{person}{Ido Roll} {and} \bibinfo{person}{Ruth
  Wylie}.} \bibinfo{year}{2016}\natexlab{}.
\newblock \showarticletitle{Evolution and revolution in artificial intelligence
  in education}.
\newblock \bibinfo{journal}{\emph{International Journal of Artificial
  Intelligence in Education}} \bibinfo{volume}{26}, \bibinfo{number}{2}
  (\bibinfo{year}{2016}), \bibinfo{pages}{582--599}.
\newblock


\bibitem[Rosner et~al\mbox{.}(2016)]%
        {rosner2016out}
\bibfield{author}{\bibinfo{person}{Daniela~K Rosner}, \bibinfo{person}{Saba
  Kawas}, \bibinfo{person}{Wenqi Li}, \bibinfo{person}{Nicole Tilly}, {and}
  \bibinfo{person}{Yi-Chen Sung}.} \bibinfo{year}{2016}\natexlab{}.
\newblock \showarticletitle{Out of time, out of place: Reflections on design
  workshops as a research method}. In \bibinfo{booktitle}{\emph{Proceedings of
  the 19th ACM Conference on Computer-Supported Cooperative Work \& Social
  Computing}}. \bibinfo{pages}{1131--1141}.
\newblock


\bibitem[Saltz and Grady(2017)]%
        {saltz2017ambiguity}
\bibfield{author}{\bibinfo{person}{Jeffrey~S Saltz} {and}
  \bibinfo{person}{Nancy~W Grady}.} \bibinfo{year}{2017}\natexlab{}.
\newblock \showarticletitle{The ambiguity of data science team roles and the
  need for a data science workforce framework}. In
  \bibinfo{booktitle}{\emph{2017 IEEE international conference on big data (Big
  Data)}}. IEEE, \bibinfo{pages}{2355--2361}.
\newblock


\bibitem[Sambasivan et~al\mbox{.}(2021)]%
        {10.1145/3411764.3445518}
\bibfield{author}{\bibinfo{person}{Nithya Sambasivan}, \bibinfo{person}{Shivani
  Kapania}, \bibinfo{person}{Hannah Highfill}, \bibinfo{person}{Diana Akrong},
  \bibinfo{person}{Praveen Paritosh}, {and} \bibinfo{person}{Lora~M Aroyo}.}
  \bibinfo{year}{2021}\natexlab{}.
\newblock \showarticletitle{“Everyone Wants to Do the Model Work, Not the
  Data Work”: Data Cascades in High-Stakes AI}. In
  \bibinfo{booktitle}{\emph{Proceedings of the 2021 CHI Conference on Human
  Factors in Computing Systems}} (Yokohama, Japan) \emph{(\bibinfo{series}{CHI
  '21})}. \bibinfo{publisher}{Association for Computing Machinery},
  \bibinfo{address}{New York, NY, USA}, Article \bibinfo{articleno}{39},
  \bibinfo{numpages}{15}~pages.
\newblock
\showISBNx{9781450380966}
\urldef\tempurl%
\url{https://doi.org/10.1145/3411764.3445518}
\showDOI{\tempurl}


\bibitem[Sanders and Stappers(2008)]%
        {doi:10.1080/15710880701875068}
\bibfield{author}{\bibinfo{person}{Elizabeth B.-N. Sanders} {and}
  \bibinfo{person}{Pieter~Jan Stappers}.} \bibinfo{year}{2008}\natexlab{}.
\newblock \showarticletitle{Co-creation and the new landscapes of design}.
\newblock \bibinfo{journal}{\emph{CoDesign}} \bibinfo{volume}{4},
  \bibinfo{number}{1} (\bibinfo{year}{2008}), \bibinfo{pages}{5--18}.
\newblock
\urldef\tempurl%
\url{https://doi.org/10.1080/15710880701875068}
\showDOI{\tempurl}
\showeprint{https://doi.org/10.1080/15710880701875068}


\bibitem[Schiff(2021)]%
        {schiff2021out}
\bibfield{author}{\bibinfo{person}{Daniel Schiff}.}
  \bibinfo{year}{2021}\natexlab{}.
\newblock \showarticletitle{Out of the laboratory and into the classroom: the
  future of artificial intelligence in education}.
\newblock \bibinfo{journal}{\emph{AI \& society}} \bibinfo{volume}{36},
  \bibinfo{number}{1} (\bibinfo{year}{2021}), \bibinfo{pages}{331--348}.
\newblock


\bibitem[Schiff(2022)]%
        {schiff2022education}
\bibfield{author}{\bibinfo{person}{Daniel Schiff}.}
  \bibinfo{year}{2022}\natexlab{}.
\newblock \showarticletitle{Education for AI, not AI for Education: the role of
  education and ethics in national AI policy strategies}.
\newblock \bibinfo{journal}{\emph{International Journal of Artificial
  Intelligence in Education}} \bibinfo{volume}{32}, \bibinfo{number}{3}
  (\bibinfo{year}{2022}), \bibinfo{pages}{527--563}.
\newblock


\bibitem[Sculley et~al\mbox{.}(2015)]%
        {sculley2015hidden}
\bibfield{author}{\bibinfo{person}{David Sculley}, \bibinfo{person}{Gary Holt},
  \bibinfo{person}{Daniel Golovin}, \bibinfo{person}{Eugene Davydov},
  \bibinfo{person}{Todd Phillips}, \bibinfo{person}{Dietmar Ebner},
  \bibinfo{person}{Vinay Chaudhary}, \bibinfo{person}{Michael Young},
  \bibinfo{person}{Jean-Francois Crespo}, {and} \bibinfo{person}{Dan
  Dennison}.} \bibinfo{year}{2015}\natexlab{}.
\newblock \showarticletitle{Hidden technical debt in machine learning systems}.
\newblock \bibinfo{journal}{\emph{Advances in neural information processing
  systems}}  \bibinfo{volume}{28} (\bibinfo{year}{2015}).
\newblock


\bibitem[Shi et~al\mbox{.}(2020)]%
        {shi2020artificial}
\bibfield{author}{\bibinfo{person}{Zheyuan~Ryan Shi}, \bibinfo{person}{Claire
  Wang}, {and} \bibinfo{person}{Fei Fang}.} \bibinfo{year}{2020}\natexlab{}.
\newblock \showarticletitle{Artificial intelligence for social good: A survey}.
\newblock \bibinfo{journal}{\emph{arXiv preprint arXiv:2001.01818}}
  (\bibinfo{year}{2020}).
\newblock


\bibitem[Sloane et~al\mbox{.}(2020)]%
        {sloane2020participation}
\bibfield{author}{\bibinfo{person}{Mona Sloane}, \bibinfo{person}{Emanuel
  Moss}, \bibinfo{person}{Olaitan Awomolo}, {and} \bibinfo{person}{Laura
  Forlano}.} \bibinfo{year}{2020}\natexlab{}.
\newblock \showarticletitle{Participation is not a design fix for machine
  learning}.
\newblock \bibinfo{journal}{\emph{arXiv preprint arXiv:2007.02423}}
  (\bibinfo{year}{2020}).
\newblock


\bibitem[Steen(2013)]%
        {10.1162/DESI_a_00207}
\bibfield{author}{\bibinfo{person}{Marc Steen}.}
  \bibinfo{year}{2013}\natexlab{}.
\newblock \showarticletitle{{Co-Design as a Process of Joint Inquiry and
  Imagination}}.
\newblock \bibinfo{journal}{\emph{Design Issues}} \bibinfo{volume}{29},
  \bibinfo{number}{2} (\bibinfo{date}{04} \bibinfo{year}{2013}),
  \bibinfo{pages}{16--28}.
\newblock
\showISSN{0747-9360}
\urldef\tempurl%
\url{https://doi.org/10.1162/DESI_a_00207}
\showDOI{\tempurl}
\showeprint{https://direct.mit.edu/desi/article-pdf/29/2/16/1715163/desi\_a\_00207.pdf}


\bibitem[Strauss and Corbin(1990)]%
        {strauss1990basics}
\bibfield{author}{\bibinfo{person}{Anselm Strauss} {and}
  \bibinfo{person}{Juliet Corbin}.} \bibinfo{year}{1990}\natexlab{}.
\newblock \bibinfo{booktitle}{\emph{Basics of qualitative research}}.
\newblock \bibinfo{publisher}{Sage publications}.
\newblock


\bibitem[Subramonyam et~al\mbox{.}(2022a)]%
        {10.1145/3491102.3517537}
\bibfield{author}{\bibinfo{person}{Hariharan Subramonyam},
  \bibinfo{person}{Jane Im}, \bibinfo{person}{Colleen Seifert}, {and}
  \bibinfo{person}{Eytan Adar}.} \bibinfo{year}{2022}\natexlab{a}.
\newblock \showarticletitle{Solving Separation-of-Concerns Problems in
  Collaborative Design of Human-AI Systems through Leaky Abstractions}. In
  \bibinfo{booktitle}{\emph{Proceedings of the 2022 CHI Conference on Human
  Factors in Computing Systems}} (New Orleans, LA, USA)
  \emph{(\bibinfo{series}{CHI '22})}. \bibinfo{publisher}{Association for
  Computing Machinery}, \bibinfo{address}{New York, NY, USA}, Article
  \bibinfo{articleno}{481}, \bibinfo{numpages}{21}~pages.
\newblock
\showISBNx{9781450391573}
\urldef\tempurl%
\url{https://doi.org/10.1145/3491102.3517537}
\showDOI{\tempurl}


\bibitem[Subramonyam et~al\mbox{.}(2022b)]%
        {subramonyam2022solving}
\bibfield{author}{\bibinfo{person}{Hariharan Subramonyam},
  \bibinfo{person}{Jane Im}, \bibinfo{person}{Colleen Seifert}, {and}
  \bibinfo{person}{Eytan Adar}.} \bibinfo{year}{2022}\natexlab{b}.
\newblock \showarticletitle{Solving Separation-of-Concerns Problems in
  Collaborative Design of Human-AI Systems through Leaky Abstractions}. In
  \bibinfo{booktitle}{\emph{CHI Conference on Human Factors in Computing
  Systems}}. \bibinfo{pages}{1--21}.
\newblock


\bibitem[Subramonyam et~al\mbox{.}(2021a)]%
        {subramonyam2021towards}
\bibfield{author}{\bibinfo{person}{Hariharan Subramonyam},
  \bibinfo{person}{Colleen Seifert}, {and} \bibinfo{person}{Eytan Adar}.}
  \bibinfo{year}{2021}\natexlab{a}.
\newblock \showarticletitle{Towards a process model for co-creating AI
  experiences}. In \bibinfo{booktitle}{\emph{Designing Interactive Systems
  Conference 2021}}. \bibinfo{pages}{1529--1543}.
\newblock


\bibitem[Subramonyam et~al\mbox{.}(2021b)]%
        {subramonyam2021can}
\bibfield{author}{\bibinfo{person}{Hariharan Subramonyam},
  \bibinfo{person}{Colleen Seifert}, {and} \bibinfo{person}{MI~Eytan Adar}.}
  \bibinfo{year}{2021}\natexlab{b}.
\newblock \showarticletitle{How Can Human-Centered Design Shape Data-Centric
  AI?}. In \bibinfo{booktitle}{\emph{NeurIPS Data-Centric AI Workshop.
  Retrieved from https://haridecoded. com/resources/AIX\_NeurIPS\_2021. pdf}}.
\newblock


\bibitem[Taylor et~al\mbox{.}(2015)]%
        {taylor2015data}
\bibfield{author}{\bibinfo{person}{Alex~S Taylor}, \bibinfo{person}{Si{\^a}n
  Lindley}, \bibinfo{person}{Tim Regan}, \bibinfo{person}{David Sweeney},
  \bibinfo{person}{Vasillis Vlachokyriakos}, \bibinfo{person}{Lillie Grainger},
  {and} \bibinfo{person}{Jessica Lingel}.} \bibinfo{year}{2015}\natexlab{}.
\newblock \showarticletitle{Data-in-place: Thinking through the relations
  between data and community}. In \bibinfo{booktitle}{\emph{Proceedings of the
  33rd Annual ACM Conference on Human Factors in Computing Systems}}.
  \bibinfo{pages}{2863--2872}.
\newblock


\bibitem[Toma{\v{s}}ev et~al\mbox{.}(2020)]%
        {tomavsev2020ai}
\bibfield{author}{\bibinfo{person}{Nenad Toma{\v{s}}ev},
  \bibinfo{person}{Julien Cornebise}, \bibinfo{person}{Frank Hutter},
  \bibinfo{person}{Shakir Mohamed}, \bibinfo{person}{Angela Picciariello},
  \bibinfo{person}{Bec Connelly}, \bibinfo{person}{Danielle Belgrave},
  \bibinfo{person}{Daphne Ezer}, \bibinfo{person}{Fanny Cachat van~der Haert},
  \bibinfo{person}{Frank Mugisha}, {et~al\mbox{.}}}
  \bibinfo{year}{2020}\natexlab{}.
\newblock \showarticletitle{AI for social good: unlocking the opportunity for
  positive impact}.
\newblock \bibinfo{journal}{\emph{Nature Communications}} \bibinfo{volume}{11},
  \bibinfo{number}{1} (\bibinfo{year}{2020}), \bibinfo{pages}{1--6}.
\newblock


\bibitem[Veale et~al\mbox{.}(2018)]%
        {veale2018fairness}
\bibfield{author}{\bibinfo{person}{Michael Veale}, \bibinfo{person}{Max
  Van~Kleek}, {and} \bibinfo{person}{Reuben Binns}.}
  \bibinfo{year}{2018}\natexlab{}.
\newblock \showarticletitle{Fairness and accountability design needs for
  algorithmic support in high-stakes public sector decision-making}. In
  \bibinfo{booktitle}{\emph{Proceedings of the 2018 chi conference on human
  factors in computing systems}}. \bibinfo{pages}{1--14}.
\newblock


\bibitem[Vertesi and Dourish(2011)]%
        {vertesi2011value}
\bibfield{author}{\bibinfo{person}{Janet Vertesi} {and} \bibinfo{person}{Paul
  Dourish}.} \bibinfo{year}{2011}\natexlab{}.
\newblock \showarticletitle{The value of data: considering the context of
  production in data economies}. In \bibinfo{booktitle}{\emph{Proceedings of
  the ACM 2011 conference on Computer supported cooperative work}}.
  \bibinfo{pages}{533--542}.
\newblock


\bibitem[Wang et~al\mbox{.}(2022)]%
        {wang2022documentation}
\bibfield{author}{\bibinfo{person}{April~Yi Wang}, \bibinfo{person}{Dakuo
  Wang}, \bibinfo{person}{Jaimie Drozdal}, \bibinfo{person}{Michael Muller},
  \bibinfo{person}{Soya Park}, \bibinfo{person}{Justin~D Weisz},
  \bibinfo{person}{Xuye Liu}, \bibinfo{person}{Lingfei Wu}, {and}
  \bibinfo{person}{Casey Dugan}.} \bibinfo{year}{2022}\natexlab{}.
\newblock \showarticletitle{Documentation Matters: Human-Centered AI System to
  Assist Data Science Code Documentation in Computational Notebooks}.
\newblock \bibinfo{journal}{\emph{ACM Transactions on Computer-Human
  Interaction}} \bibinfo{volume}{29}, \bibinfo{number}{2}
  (\bibinfo{year}{2022}), \bibinfo{pages}{1--33}.
\newblock


\bibitem[Weber et~al\mbox{.}(2022)]%
        {weber2022organizational}
\bibfield{author}{\bibinfo{person}{Michael Weber}, \bibinfo{person}{Martin
  Engert}, \bibinfo{person}{Norman Schaffer}, \bibinfo{person}{J{\"o}rg
  Weking}, {and} \bibinfo{person}{Helmut Krcmar}.}
  \bibinfo{year}{2022}\natexlab{}.
\newblock \showarticletitle{Organizational capabilities for ai
  implementation—coping with inscrutability and data dependency in ai}.
\newblock \bibinfo{journal}{\emph{Information Systems Frontiers}}
  (\bibinfo{year}{2022}), \bibinfo{pages}{1--21}.
\newblock


\bibitem[Whang et~al\mbox{.}(2023)]%
        {whang2023data}
\bibfield{author}{\bibinfo{person}{Steven~Euijong Whang}, \bibinfo{person}{Yuji
  Roh}, \bibinfo{person}{Hwanjun Song}, {and} \bibinfo{person}{Jae-Gil Lee}.}
  \bibinfo{year}{2023}\natexlab{}.
\newblock \showarticletitle{Data collection and quality challenges in deep
  learning: A data-centric ai perspective}.
\newblock \bibinfo{journal}{\emph{The VLDB Journal}} (\bibinfo{year}{2023}),
  \bibinfo{pages}{1--23}.
\newblock


\bibitem[Whittaker et~al\mbox{.}(2018)]%
        {whittaker2018ai}
\bibfield{author}{\bibinfo{person}{Meredith Whittaker}, \bibinfo{person}{Kate
  Crawford}, \bibinfo{person}{Roel Dobbe}, \bibinfo{person}{Genevieve Fried},
  \bibinfo{person}{Elizabeth Kaziunas}, \bibinfo{person}{Varoon Mathur},
  \bibinfo{person}{Sarah~Mysers West}, \bibinfo{person}{Rashida Richardson},
  \bibinfo{person}{Jason Schultz}, {and} \bibinfo{person}{Oscar Schwartz}.}
  \bibinfo{year}{2018}\natexlab{}.
\newblock \bibinfo{booktitle}{\emph{AI now report 2018}}.
\newblock \bibinfo{publisher}{AI Now Institute at New York University New
  York}.
\newblock


\bibitem[Williams and Begg(1993)]%
        {williams1993translation}
\bibfield{author}{\bibinfo{person}{Marian~G Williams} {and}
  \bibinfo{person}{Vivienne Begg}.} \bibinfo{year}{1993}\natexlab{}.
\newblock \showarticletitle{Translation between software designers and users}.
\newblock \bibinfo{journal}{\emph{Commun. ACM}} \bibinfo{volume}{36},
  \bibinfo{number}{6} (\bibinfo{year}{1993}), \bibinfo{pages}{102--103}.
\newblock


\bibitem[Zhang et~al\mbox{.}(2020)]%
        {zhang2020data}
\bibfield{author}{\bibinfo{person}{Amy~X Zhang}, \bibinfo{person}{Michael
  Muller}, {and} \bibinfo{person}{Dakuo Wang}.}
  \bibinfo{year}{2020}\natexlab{}.
\newblock \showarticletitle{How do data science workers collaborate? roles,
  workflows, and tools}.
\newblock \bibinfo{journal}{\emph{Proceedings of the ACM on Human-Computer
  Interaction}} \bibinfo{volume}{4}, \bibinfo{number}{CSCW1}
  (\bibinfo{year}{2020}), \bibinfo{pages}{1--23}.
\newblock


\bibitem[Zhang et~al\mbox{.}(2022)]%
        {zhang2022storybuddy}
\bibfield{author}{\bibinfo{person}{Zheng Zhang}, \bibinfo{person}{Ying Xu},
  \bibinfo{person}{Yanhao Wang}, \bibinfo{person}{Bingsheng Yao},
  \bibinfo{person}{Daniel Ritchie}, \bibinfo{person}{Tongshuang Wu},
  \bibinfo{person}{Mo Yu}, \bibinfo{person}{Dakuo Wang}, {and}
  \bibinfo{person}{Toby Jia-Jun Li}.} \bibinfo{year}{2022}\natexlab{}.
\newblock \showarticletitle{StoryBuddy: A Human-AI Collaborative Chatbot for
  Parent-Child Interactive Storytelling with Flexible Parental Involvement}. In
  \bibinfo{booktitle}{\emph{CHI Conference on Human Factors in Computing
  Systems}}. \bibinfo{pages}{1--21}.
\newblock


\bibitem[Zhou et~al\mbox{.}(2021)]%
        {zhou2021investigating}
\bibfield{author}{\bibinfo{person}{Qi Zhou}, \bibinfo{person}{Wannapon
  Suraworachet}, \bibinfo{person}{Stanislav Pozdniakov},
  \bibinfo{person}{Roberto Martinez-Maldonado}, \bibinfo{person}{Tom
  Bartindale}, \bibinfo{person}{Peter Chen}, \bibinfo{person}{Dan Richardson},
  {and} \bibinfo{person}{Mutlu Cukurova}.} \bibinfo{year}{2021}\natexlab{}.
\newblock \showarticletitle{Investigating students’ experiences with
  collaboration analytics for remote group meetings}. In
  \bibinfo{booktitle}{\emph{International Conference on Artificial Intelligence
  in Education}}. Springer, \bibinfo{pages}{472--485}.
\newblock


\end{thebibliography}

\end{document}